\documentclass[review]{elsarticle}
\usepackage[mathscr]{euscript}
\usepackage[utf8]{inputenc} 
\usepackage[T1]{fontenc}    
\usepackage{url}            
\usepackage{booktabs}       
\usepackage{float}
\floatstyle{plaintop}
\restylefloat{table}
\usepackage{multirow}
\usepackage{multicol}
\usepackage{amsfonts}       
\usepackage{nicefrac}       
\usepackage{microtype}      
\usepackage{natbib}
\usepackage{amsmath}
\usepackage{color, soul} 
\setulcolor{red}
\sethlcolor{blue}
\renewcommand\hl[1]{#1} 

\usepackage{breqn}
\usepackage{lineno}
\usepackage[margin=3cm]{geometry}
\usepackage{textcomp}
\usepackage{gensymb}
\usepackage{diagbox}
\usepackage{rotating}
\usepackage{enumerate}
\usepackage{caption} 
\usepackage{amssymb}
\usepackage{bm}
\usepackage{siunitx}
\newcommand{\href}[1]{#1} 
\usepackage[pdftex,pagebackref=false]{hyperref} 
\hypersetup{
    pdfnewwindow=true,      
    pdfauthor=author,
    colorlinks=true,        
    linkcolor=black,         
    citecolor=blue,        
    filecolor=black,      
    urlcolor=black           
}

\journal{ }

\begin{document}

\begin{frontmatter}

\title{Architectural bone parameters and the relationship to titanium lattice design for powder bed fusion additive manufacturing}

\author[mymainaddress]{Martine McGregor}
\author[mymainaddress]{Sagar Patel}
\author[mymainaddress]{Stewart McLachlin}
\author[mymainaddress]{Mihaela Vlasea \corref{mycorrespondingauthor}}
\cortext[mycorrespondingauthor]{Corresponding author}
\ead{mihaela.vlasea@uwaterloo.ca}

\address[mymainaddress]{University of Waterloo, Department of Mechanical and Mechatronics Engineering, Waterloo, ON N2L 3G1, Canada}

\begin{abstract}
Additive manufacturing (AM) of titanium (Ti) and Ti-6Al-4V lattices has been proposed for bone implants and augmentation devices. Ti and Ti-6Al-4V have favourable biocompatibility, corrosion resistance and fatigue strength for bone applications; yet, the optimal parameters for Ti-6Al-4V lattice designs corresponding to the natural micro- and meso-scale architecture of human trabecular and cortical bone are not well understood.  A comprehensive review was completed to compare the natural lattice architecture properties in human bone to Ti and Ti-6Al-4V lattice structures for bone replacement and repair. Ti and Ti-6Al-4V lattice porosity has varied from 15\% to 97\% with most studies reporting a porosity between 50-70\%. Cortical bone is roughly 5-15\% porous and lattices with 50-70\% porosity are able to achieve comparable stiffness, compressive strength, and yield strength. Trabecular bone has a reported porosity range from 70-90\%, with trabecular thickness varying from 120-200 \textmu m. \hl{Existing powder bed fusion technologies have produced strut and wall thicknesses ranging from 200-1669 \textmu m}. This suggests limited overlap between current AM of Ti and Ti-6Al-4V lattice structures and trabecular bone architecture, indicating that replicating natural trabecular bone parameters with latticing is prohibitively challenging. This review contributes to the body of knowledge by identifying the correspondence of Ti and Ti-6Al-4V lattices to the natural parameters of bone microarchitectures, and provides further guidance on the design and AM recommendations towards addressing recognized performance gaps with powder bed fusion technologies.  
\end{abstract}


\end{frontmatter}


\section{Introduction}
The use of Ti-6Al-4V is well established in the medical device industry \cite{oshida2010bioscience}. Titanium and titanium alloys are ideal for replacing hard tissues, such as bone, due to their biocompatibility and excellent strength-to-weight ratio \cite{song2017design, lin2013additive}. Ti-6Al-4V also exhibits excellent fatigue strength and corrosion resistance allowing implants to withstand the cyclic loading in high ion environments present in vivo during activities of daily living \cite{ong2014introduction}. These properties of titanium and Ti-6Al-4V have led such alloys to become some of the most widely used metal materials for bone replacement and fixation procedures in the orthopaedic industry.

Despite their widespread use in the medical device industry for the replacement and repair of bone, pure titanium and Ti-6Al-4V do not have material properties similar to those of bone. Ti-6Al-4V is roughly twice as stiff as human cortical bone with up to seven times the compressive strength \cite{emmelmann2011laser, du2019beautiful, murr2011microstructure}. Bone is mechanoresponsive and requires regular loading in order to proliferate new bone. Implanting a stiffer material like titanium adjacent to bone can cause stress shielding of bone, resorption of the surrounding bone tissue which can lead to implant loosening and failure \cite{tonino1976protection}. Therefore, it is imperative that designs of titanium and titanium alloy implants are tailored to more closely match the natural mechanical response of bone tissue.

One approach to reducing stress shielding near the bone-implant interface is by light-weighting implants through latticing in an effort to reduce mechanical properties from those of the bulk modulus of the material. Additive manufacturing allows for unique design approaches thereby allowing for control of mechanical properties, while minimizing weight through unique geometries and graded material properties. Additive manufacturing technologies are of particular use in the \hl{orthopaedic} device industry where implants are made for specific applications in which weight and mechanical properties are integral to implant function. \hl{Some examples of orthopaedic applications for latticed titanium structures include light-weighting femoral stems for total hip replacement, optimized lattice structure for spinal interbody fusion cages and bone repair scaffolds following tumour resection procedures} \cite{arabnejad2017fully, mobbs2017utility,kim2017sacral, choy2017reconstruction, phan2016application, xu2016reconstruction}.

Production of Ti-6Al-4V lattice structures through additive manufacturing for biomedical applications was thoroughly reviewed by Tan et al. in 2017. They concluded that optimal guidelines for lattice design in biological environments has not yet been established and that the field is rapidly and continually evolving \cite{tan2017metallic}. Since the review by Tan et al. in 2017 was completed, many additional studies have successfully progressed toward this goal by manipulating lattice parameters such as porosity, pore size, and strut thickness in order to produce unit cells that more closely exhibit stiffness, compressive strength and fatigue strength of human bone. Lattice parameters vary widely across the literature and while general recommendations have been made, there remain questions as to which lattice design produces the optimal structure for implant fixation and reduction of stress shielding through mechanical property optimization.

A consideration that is notably overlooked in developing lattice structures for bone replacement is how to best model the architectural parameters of human bone. Therefore, a comprehensive literature review is needed to examine the parameters of bone micro-architecture in correspondence with lattice design parameters used in additive manufacturing. A review of existing additive manufacturing literature was completed to determine which titanium lattice parameters have been examined for bone replacement and repair implants. Lastly, recommendations are made on how to best use this information towards the design and additive manufacturing of improved titanium and Ti-6Al-4V lattice structures for bone replacement and repair.

\section{Review of human bone properties relevant to lattice design}
\subsection{Review of human bone function}
\begin{figure}[htbp]
    \centering
    \captionsetup{justification=centering}
    \includegraphics[width=15cm,keepaspectratio]{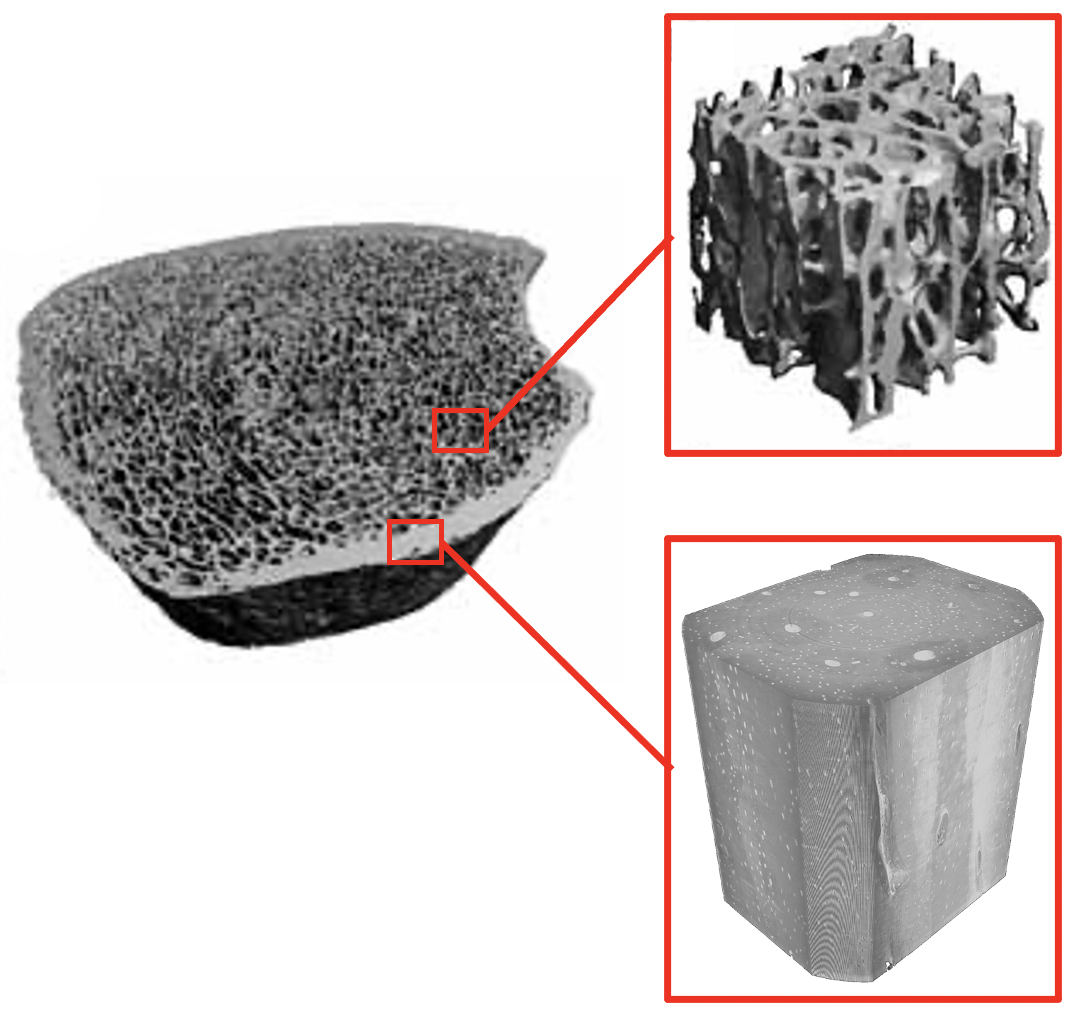}
    \caption{Human bone can be categorized into two main types: cortical and trabecular. Cortical bone is the stiffer, more dense bone which encapsulates short bones, the ends of long bones and comprises the shaft of long bones. Trabecular bone is the lattice like structure which makes up the majority of short bone structure as well as the ends of long bones. This figure was adapted from the \textmu CT work of Lui et al. and Gauthier et al. \cite{liu2010accuracy, gauthier20193d}}
    \label{fig:bonetypes}
\end{figure}

Bone is a tough, elastic tissue that gives structural support to the human body. As a living tissue, bone adapts to its environment and loading conditions through the breakdown of existing bone by osteoclasts and the proliferation of new bone by osteoblasts. This process leads bone to exhibit mechanoresponsive behaviour, wherein the more it is loaded, the thicker and denser it becomes. Bone can be categorized into two main types: cortical bone and trabecular bone, see Figure \ref{fig:bonetypes}. Different bones in the human body consist of different amounts and configurations of cortical and trabecular bone. Long bones, such as the femur, tibia, radius, and humerus, have a long shaft made up of primarily cortical bone. The articulating ends of long bones are made of primarily trabecular bone contained by a thin shell of cortical bone. Short bones, such as vertebrae, carpals, and tarsals, are made primarily of trabecular bone with a thin cortical shell making them strong and compact. Cortical bone density and trabecular bone density are important indicators of bone strength, and decrease with age, for instance by approximately 0.41\% and 0.65\% per year, respectively, for women ages 70 to 87 years old \cite{boonen1997factors}. However, studies have demonstrated that trabecular bone has a higher correlation to bone strength than cortical bone \cite{spadaro1994cortical}.

\subsection{Trabecular bone}
Trabecular bone, or cancellous bone, is a lattice-like structure that allows bone to maintain its strength, while being relatively lightweight. Trabecular bone is primarily located at the articulating ends of long bones and in the body of short bones, which allows for improved load transfer through these structures. The exact configuration of trabecular bone micro-structure is still being discovered. The most established theory characterizes the trabeculae, or struts, by their thickness, spacing, number and spatial configuration. Measurement techniques for these parameters are well established and widely published to comparatively describe trabecular bone quality, with more and thicker trabeculae suggesting better bone health \cite{majumdar1997correlation}. However, emerging and more advanced geometric models consider the shape of trabecular surface and the rod vs plate-like structures of trabecular architecture in greater detail \cite{callens2020local,van2006specimen}.

The integrity of trabecular bone microstructure is often used as an indicator for overall bone health. At any given age, men have a higher trabecular bone mass than women, but decreases in trabecular bone density due to aging are similar in women and men \cite{chen2013age,riggs2004population,khosla2006effects,nicks2012relationship,macdonald2011age}. Age-related changes cause buckling of trabeculae due to a decrease in the number and thickness of trabeculae, and an increase in trabeculae length \cite{majumdar1997correlation}. Collectively, these changes result in a decreased trabecular density; however, the pathology of this reduction differs in men and women \cite{majumdar1997correlation}. In women, the decrease in bone volume occurs primarily due to a decrease in the number of trabeculae, whereas in men, it may be predominantly attributed to the thinning of trabeculae \cite{chen2013age,aaron1987microanatomy}. The decrease in trabeculae in women is related to menopause, where less estrogen is produced, which increases bone reabsorption \cite{seeman2002pathogenesis}. Therefore, in making design considerations for orthopaedic medical devices, the patient population, age and sex should be carefully considered. Changes may be made to lattices designs to model the reduction in bone density with age. Lattice structures with fewer shorter features may better represent aging female bone and lengthening features may better model aging male bone. Lower density lattice designs should also be considered for older adults and post-menopausal women.

\subsubsection{Micro-architecture of trabecular bone}
\begin{figure}[htbp]
    \centering
    \captionsetup{justification=centering}
    \includegraphics[width=15cm,keepaspectratio]{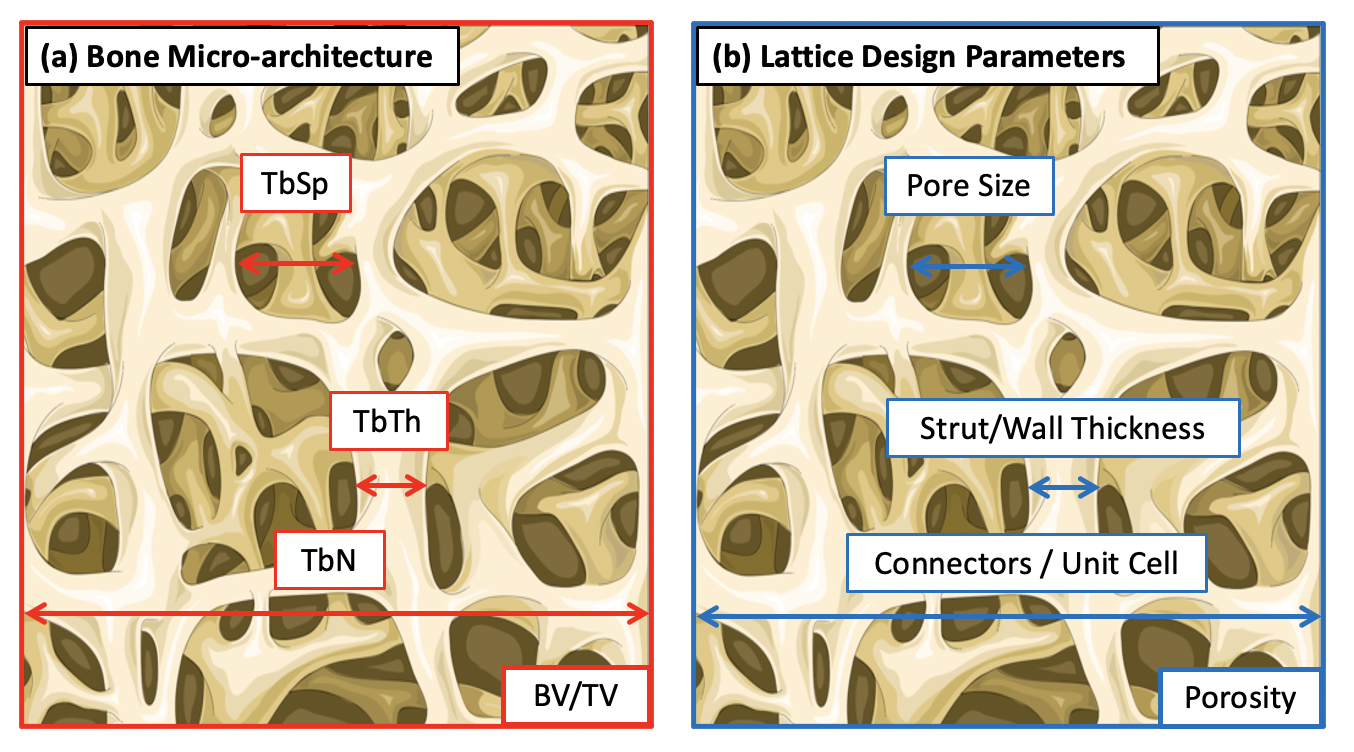}
    \caption{Commonly reported measurements of trabecular bone microstructure (red) may be used to describe lattice parameters commonly used in additive manufacturing (blue).}
    \label{fig:TrabecularBone}
\end{figure}

The micro-structure of trabecular bone can be characterized by the individual trabeculae and the spaces between them. To determine bone porosity, the volume of bone (BV) is divided by the total volume (TV) for a given sample (BV/TV). Other groups have also compared total bone surface area (BS) with respect to bone volume (BS/BV) or total specimen volume (BS/TV) as another form of trabecular quality indicator. Trabeculae are further characterized by thickness of individual vertical trabeculae (TbTh), the distance between trabeculae (TbSp), and the number of trabeculae that intersect with a given two-dimensional distance metric (TbN), typically 1 cm, as shown in Figure \ref{fig:TrabecularBone}.

Methods of calculating bone architectural parameters may vary between studies; however, the technique proposed by Hildebrand et al. is the most widely accepted in general literature, as shown in Table \ref{tab:table_1}. Hildebrand et al. proposed a two-dimensional (2D) plate model that allows for calculation of 2D parameters such as TbTh, TbSp and TbN \cite{hildebrand1999direct}.

\begin{table}[htbp]
    \centering
    \captionsetup{justification=centering}
    \begin{tabular}{@{}lc@{}}
       \toprule
       \textbf{Trabecular Bone Parameter} &  \textbf{Measurement/Calculation}\\
       \midrule \midrule
       Apparent bone density & BV/TV\\
       Porosity	& 1 - BV/TV\\
       Bone surface fraction & BS/BV or BS/TV\\
       Trabecular thickness	& TbTh = 2 BV/BSs\\
       Trabecular spacing & TbSp = 2 (TV-BV)/BS\\
       Trabecular number & TbN = 0.5 BS/TV\\
       \bottomrule
    \end{tabular}
    \caption{The Hildebrand et al. method for calculating two dimensional trabecular measurements from known bone volume fractions.}
    \label{tab:table_1}
\end{table}

\begin{figure}[htbp]
    \centering
    \captionsetup{justification=centering}
    \includegraphics[width=16cm,keepaspectratio]{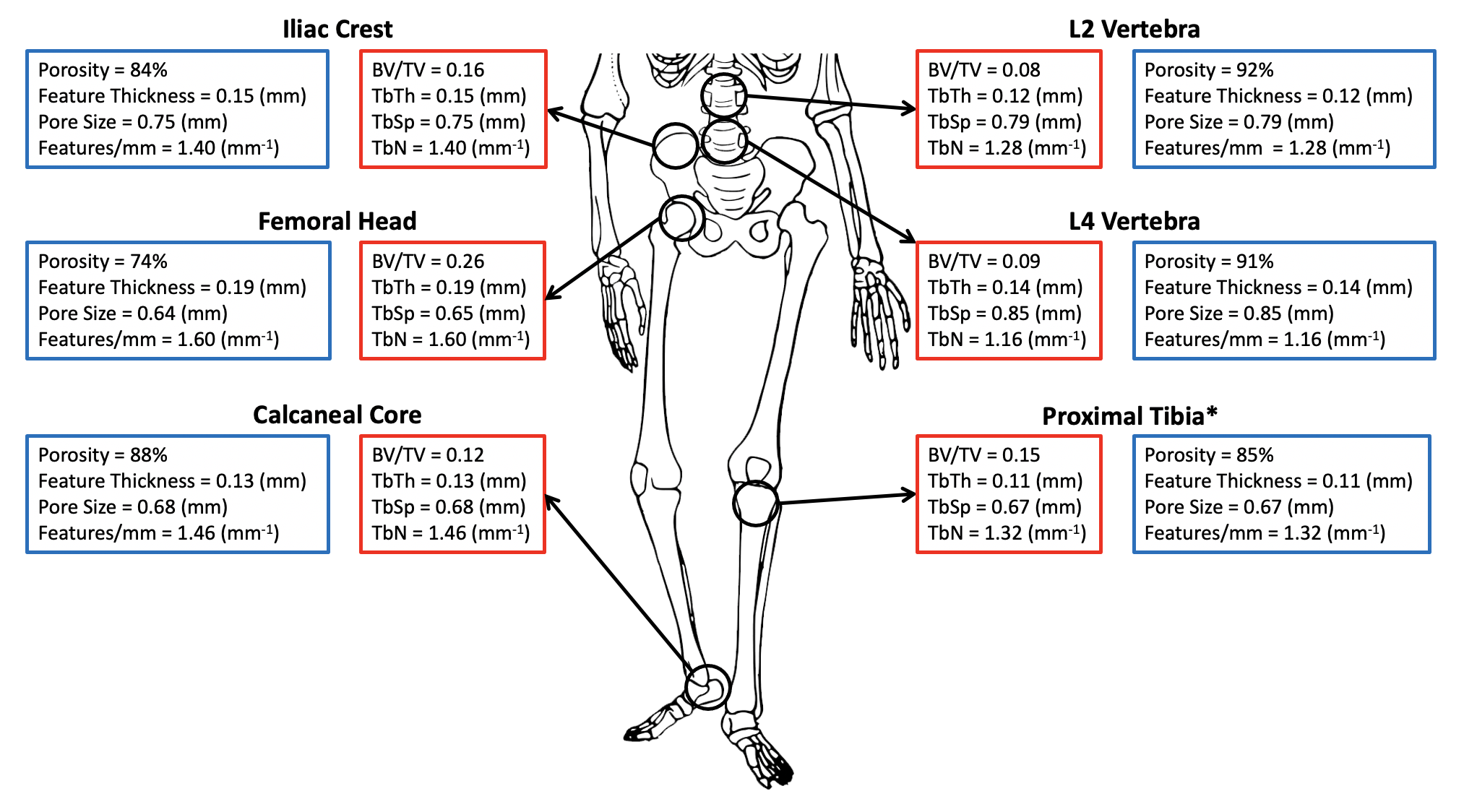}
    \caption{A comprehensive summary of two- and three-dimensional measurements of trabecular bone at the femoral head, illiac crest, calcaneal core, L2 vertebra and L4 vertebra as described by Hildebrand et al. \cite{hildebrand1999direct}. \hl{The mean data reported by Hildebrand et al. was supplemented with Thomsen et al.'s report} \cite{thomsen2005stereological} \hl{of trabecular bone parameters at the proximal tibia to form an original visual representation*}. Bone microstructure data is outlined in red and suggested translation to lattice design parameters is outlined in blue. The values listed represent the mean value of the each measurement.}
    \label{fig:skeleton}
\end{figure}

Trabecular bone density and parameters vary greatly with anatomical location, as shown in Figure \ref{fig:skeleton}. When determining design considerations for bone replacing implants, anatomical location and device function should be considered. Hildebrand et al. undertook an in-depth in vitro study comparing the micro-architecture of trabecular bone at different sites across the skeleton \cite{hildebrand1999direct}.

\subsubsection{Mechanical properties of trabecular bone}
The mechanical properties of trabecular bone vary significantly with respect to bone density. For instance, trabecular bone density predicts 81\% of axial strength variation in the human tibia \cite{munford2020mapping}. The porosity of human trabecular bone ranges from 40-95\%, dependent on skeletal location, bone region, and population parameters \cite{morgan2018bone}. \hl{Trabecular bone is anisotropic with stiffness and strength differing along the z-, x- and y-axis} \cite{liu2008complete}. Young’s modulus of trabecular bone can range from 1-5 GPa along the axis of loading and 50-700 MPa off-axis. As such, the compressive and tensile strength of trabecular bone ranges from 0.1-30 MPa and 6-8 MPa, respectively \cite{bartel2006orthopaedic,park2007biomaterials,velasco2015design}.

\subsection{Cortical bone}
Cortical bone, or compact bone, is much more dense than trabecular bone and acts as a stiff outer layer for short bones, joints in long bones and the sole composition of the hollow shafts of long bones. Cortical shells allow for a continuous surface at the joints for ligamentous and tendonous attachment. Cortical bone improves the overall fracture toughness and provides structural integrity to limbs allowing for gross movement.

\subsubsection{Micro-architecture of cortical bone}
Cortical bone is much more dense than trabecular and exhibits only 5-15\% porosity \cite{morgan2018bone}. It is well documented that cortical bone density and thickness decrease with age \cite{chen2013age}. Interestingly, micro-structural analysis has shown that the cortical bone stiffness, fatigue strength and fracture toughness decrease with age \cite{mccalden1993age}. It should also be noted that studies have indicated that the decrease in cortical bone density is significant in women, but insignificant in men \cite{chen2013age}.

\subsubsection{Mechanical properties of cortical bone}
\begin{figure}[htbp]
    \centering
    \captionsetup{justification=centering}
    \includegraphics[width=8cm,keepaspectratio]{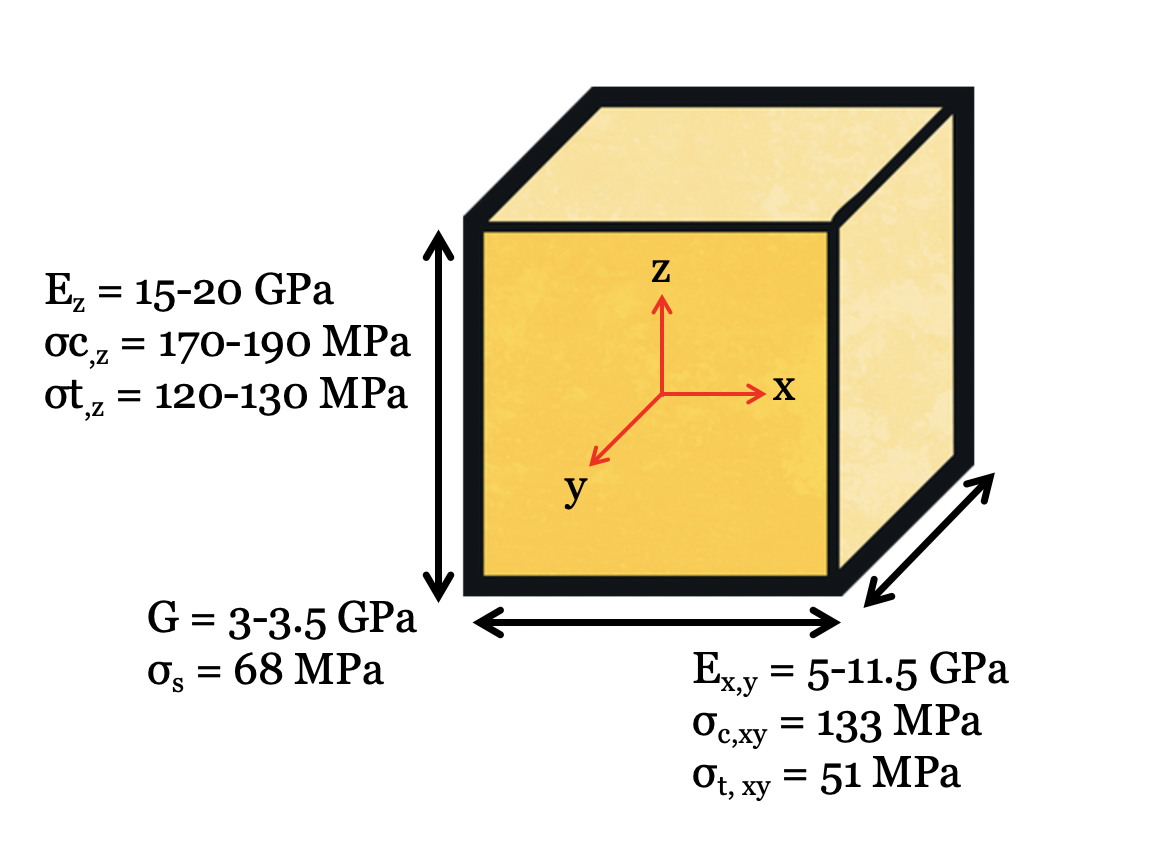}
    \caption{Cortical bone exhibits transversely isotropic mechanical properties with nearly double the stiffness occurring along the axis of loading. \cite{bartel2006orthopaedic,park2007biomaterials,burstein1976aging} Young's modulus, ultimate compressive strength, and ultimate tensile strength in the longitudinal direction are denoted E\textsubscript{z}, $\sigma_{c,z}$, and $\sigma_{t,z}$ respectively. The subscript "x,y" is given to denote mechanical properties in the transverse axes.} 
    \label{fig:transiso}
\end{figure}

Cortical bone is widely considered to be transversely isotropic, with mechanical properties along the axis of loading, or the z-axis, being significantly greater than those in the x- and y- axis. The transversely isotropic mechanical properties of human cortical bone have been collated from literature to provide a holistic visual representation, see Figure \ref{fig:transiso} \cite{bartel2006orthopaedic,park2007biomaterials,burstein1976aging}.

\section{Additively manufactured lattices for bone replacement}
\hl{The production of three-dimensional (3D) parts in AM technologies is digitally controlled, from a computer-aided design (CAD) model, to layer-by-layer customization of process parameters, to monitoring and controlling the process while parts are being manufactured. The digitization of AM opens up exciting avenues in areas of medical device design, AM process planning for device production, and advanced monitoring and control during production. The advances in CAD fuelled by the maturity of AM enable both technologies to have advanced technological readiness, enabling the regulatory pathways and component mechanical performance to be ready for wide-scale application in the medical device industry} \cite{peel2016additively, burton2018reporting}. Existing literature on additively manufactured medical devices for bone replacement or augmentation is widely spread based on target audience, research background, and study type. Research on additively manufactured medical devices can be found in surgical, tissue, biomedical engineering, additive manufacturing and/or material science journals. Collating the existing body of research into a cohesive overview proves challenging, as the focus of research covers a wide array of topics from AM process parameters, material properties, and biocompatibility, all the way to medical function and clinical outcomes. In order to review metal additive manufacturing as a technology for bone replacement an overview of metal AM and lattice design approaches has been included below.

\subsection{Review of additive manufacturing for metals: powder bed fusion}
\hl{Metal powder bed fusion (PBF) technologies such as laser powder bed fusion (LPBF) and electron beam powder bed fusion (EB-PBF) are currently the leading metal AM technologies for industrial part production for biomedical, aerospace, and automotive applications. The faster fusion of metal possible by these two technologies alongside better process automation has resulted in a forecasted dominance of LPBF and EB-PBF until 2025} \cite{Kupper2017Get}. 

\hl{LPBF (also known as selective laser melting (SLM)) and EB-PBF (also known as electron beam melting (EBM))} are AM processes in which a heat source is directed towards a powdered material to micro-weld the material together, layer-by-layer. During the LPBF process, a fine laser beam (with beam spot sizes ($\sigma$) generally between 50 to 100 \textmu m) is generally directed through a series of lenses towards an X-Y plane via scanning mirrors which direct the beam towards the build platform. In EB-PBF (beam spot sizes ($\sigma$) generally greater than 200 \textmu m) the same phenomenon is achieved through a directed high energy electron beam rather than a laser. In both cases, the build platform contains a bed of powdered material. As each layer is micro-welded together, the build platform lowers and a new layer of powder is deposited over the build platform with a blade or rake. This process is repeated layer-by-layer until the part is complete. Post-processing may be required to ensure certain material properties, part geometry and/or surface finish. 

\hl{LPBF and EB-PBF do not necessarily compete with each and rather offer different advantages for a given application, as summarized in Table}~\ref{tab:PBF_compare}. Some of the major advantages of LPBF when compared to other metal AM processes are its fine resolution \cite{sames2016metallurgy}, wide range of materials available for the technology \cite{sing2016laser}, and the potential to obtain performance superior to conventional manufacturing processes \cite{wang2018additively}. The superior resolution of LPBF when compared to direct energy deposition (DED) and EB-PBF makes it an ideal candidate for manufacturing intricate lattices used in light-weighting parts, as well as for manufacturing complex lattice structures with fine feature sizes. LPBF also has the largest range of metal material options of any metal AM technologies. One such material is Ti-6Al-4V, a widely used biomaterial in the medical device industry. \hl{There are some benefits to considering EB-PBF for the fabrication of biomedical devices: a higher build speed and reduction in residual stresses resulting in reduced part distortion. The reduction in residual stress is due to the elevated environment temperature during EB-PBF production, where the powder layer is pre-heated into a powder cake which serves as a thermal dissipation pathway, reducing the need for extensive support structures except for part substrate anchoring requirements.} As such, both LPBF and EB-PBF will be considered in this work, with a focus on lattice structure design, performance, and manufacturability for orthopaedic bone replacement and augmentation devices. 

\begin{table}[!htbp]
    \centering
    \captionsetup{justification=centering}
    \caption{\hl{Manufacturability comparisons between LPBF and EB-PBF (EBM) for fabricating titanium.}}
    \label{tab:PBF_compare}
    \begin{tabular}{@{}lcc@{}}
       \toprule
       \textbf{Metric} & \textbf{LPBF} & \textbf{EB-PBF}\\
       \midrule \midrule
        Feature resolution & $\bm{+}$ & $\bm{-}$ \\
        Manufacturing (build) speed & $\bm{-}$ & $\bm{+}$  \\
        Residual stresses \& distortion & $\bm{-}$ & $\bm{+}$  \\
        Need for support structures & $\bm{-}$ & $\bm{+}$  \\
        Feature accuracy & $\bm{+}$ & $\bm{-}$  \\
        Material choices & $\bm{+}$ & $\bm{-}$  \\
       \bottomrule
    \end{tabular}
\end{table}

\subsection{Review of lattice designs}
\begin{figure}[htbp]
    \centering
    \captionsetup{justification=centering}
    \includegraphics[width=16cm,keepaspectratio]{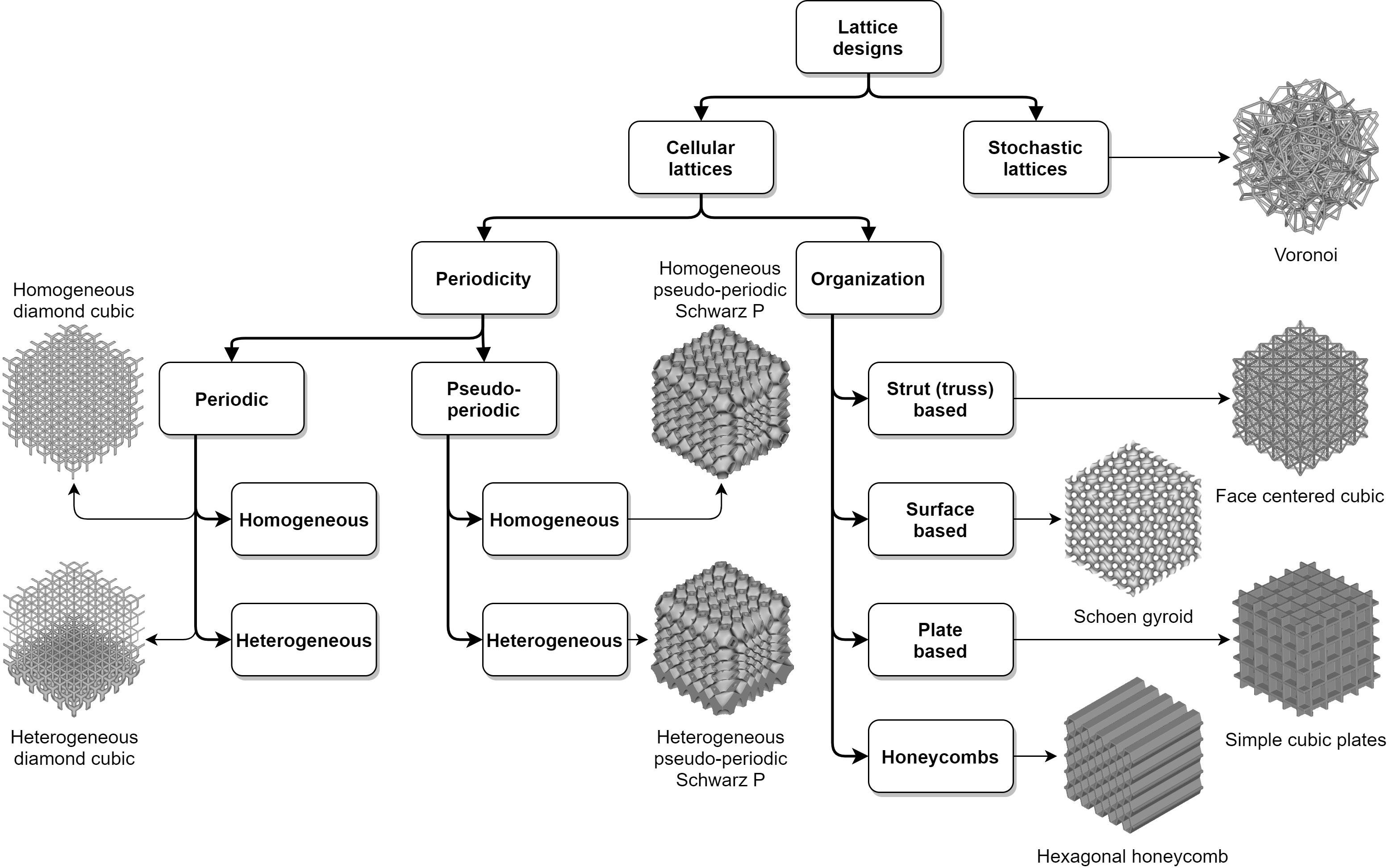}
    \caption{\hl{Classification chart for lattice designs}}
    \label{fig:lattice_types}
\end{figure}

Lattice structures are a form of hierarchical design structures used to minimize unnecessary material with respect to design function. Lattices are typically designed for a specific application or function such as reducing weight while maintaining mechanical strength, or improving energy absorption characteristics of a design component \cite{casadei2007impact, mines2013drop}. Lattice structures can be categorized into two main structure types: cellular lattices and stochastic (random) lattices \cite{gibson2014additive}. \hl{A visualization of the types of lattice structures is provided by a classification chart in Figure}~\ref{fig:lattice_types}.

\subsubsection{Cellular lattices}
The majority of lattice structures can be categorized as cellular structures that are made up of unit cells with distinct, repeatable features. \hl{Cellular lattice structures are made up of struts, walls, and/or plates that are repeatedly interconnected in 3D-space by nodes, as shown in Figure}~\ref{fig:lattice_types}.

\subsubsection*{\hl{Periodicity driven classification}}

\begin{figure}[htbp]
    \centering
    \captionsetup{justification=centering}
    \begin{tabular}{@{}lcc@{}}
        & \textbf{3D view} & \textbf{2D view} \\
         \rotatebox[origin=c]{90}{\textbf{Periodic}} & \raisebox{-.5\height}{\includegraphics[scale=0.3]{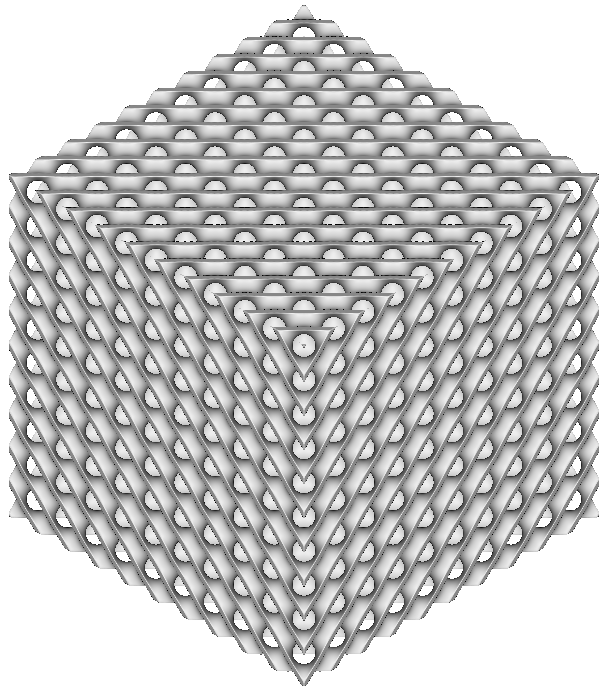}}&  \raisebox{-.5\height}{\includegraphics[scale=0.3]{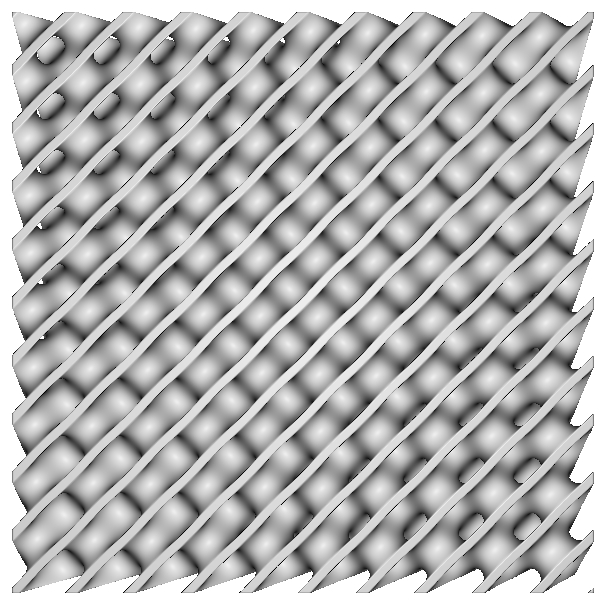}}\\
         \rotatebox[origin=c]{90}{\textbf{Pseudo-periodic}} & \raisebox{-.5\height}{\includegraphics[scale=0.3]{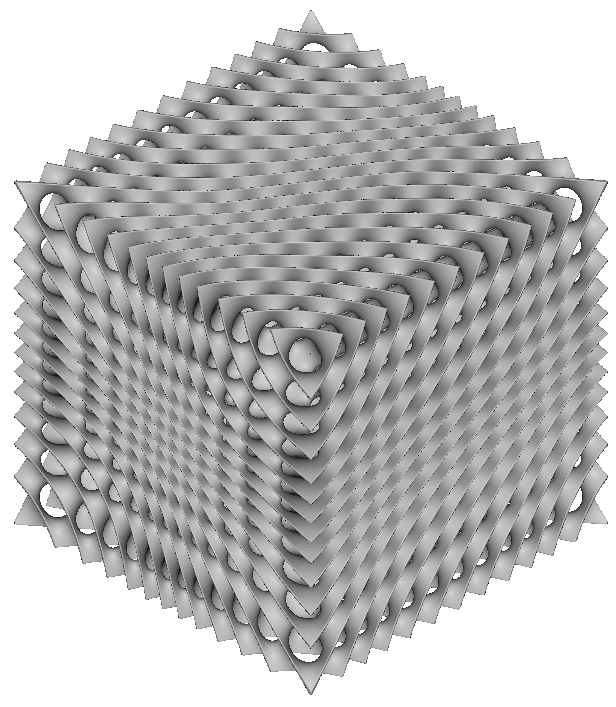}}&  \raisebox{-.5\height}{\includegraphics[scale=0.3]{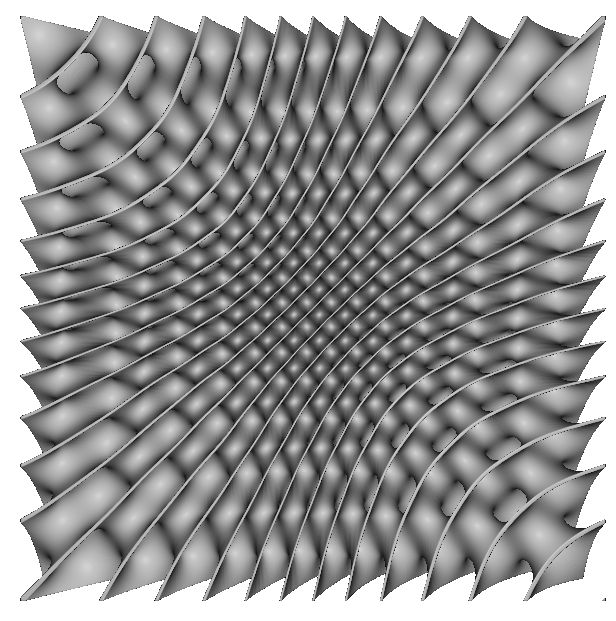}}
    \end{tabular}
    \caption{\hl{Comparative images of a periodic (top row) and pseudo-periodic (bottom row) TPMS (surface-based) Schwarz diamond (D) lattice structure.}}
    \captionsetup{justification=centering}
    \label{fig:TPMS_diamond}
\end{figure}

\begin{figure}[htbp]
    \centering
    \captionsetup{justification=centering}
    \begin{tabular}{@{}lcc@{}}
        & \textbf{3D view} & \textbf{2D view} \\
         \rotatebox[origin=c]{90}{\textbf{Homogeneous}} & \raisebox{-.5\height}{\includegraphics[scale=0.3]{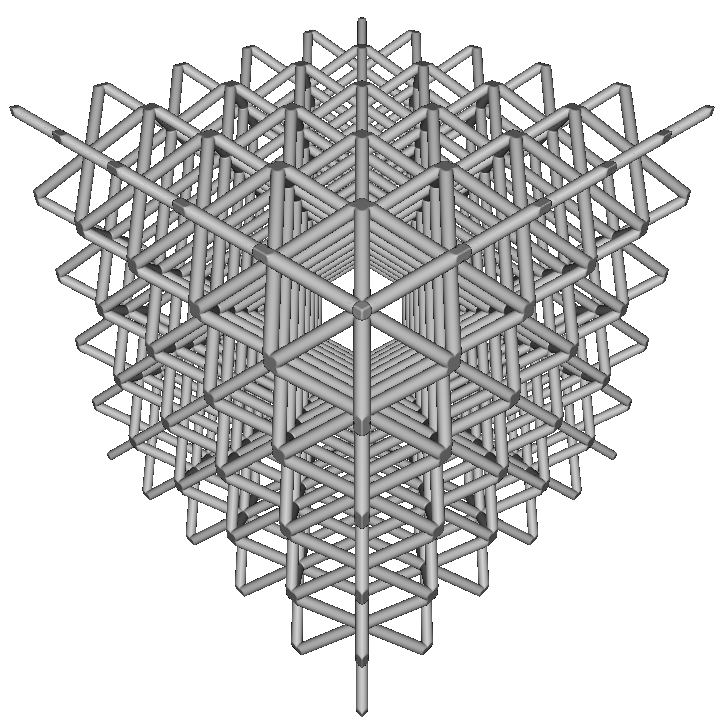}}&  \raisebox{-.5\height}{\includegraphics[scale=0.43]{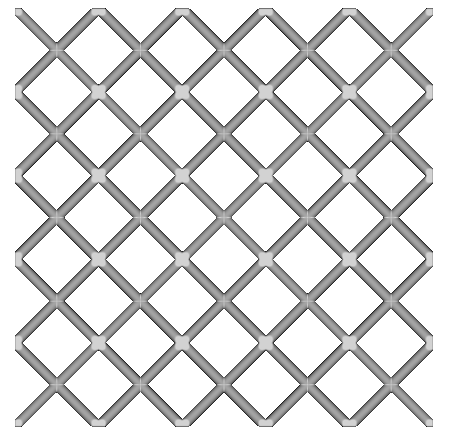}} \\
         \rotatebox[origin=c]{90}{\textbf{Heterogeneous}} & \raisebox{-.5\height}{\includegraphics[scale=0.3]{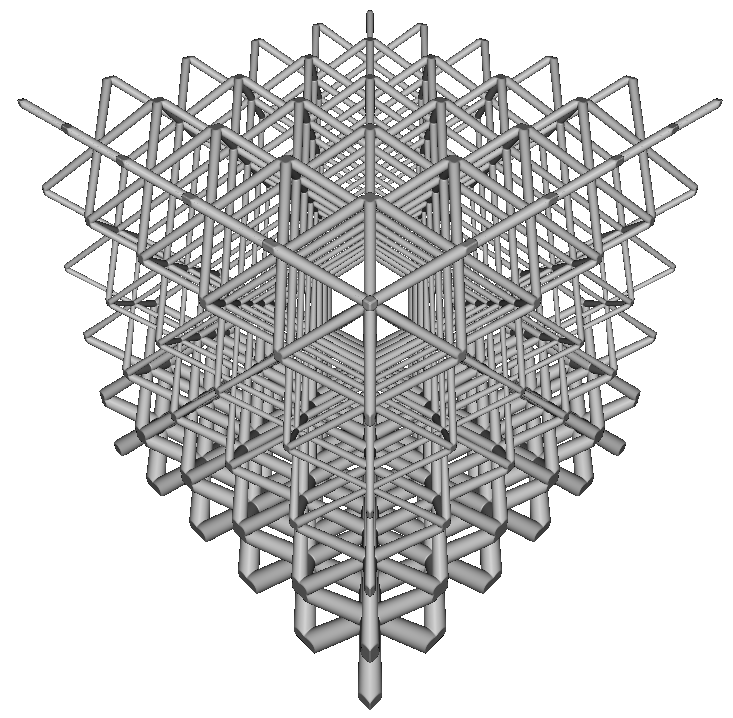}}&  \raisebox{-.5\height}{\includegraphics[scale=0.43]{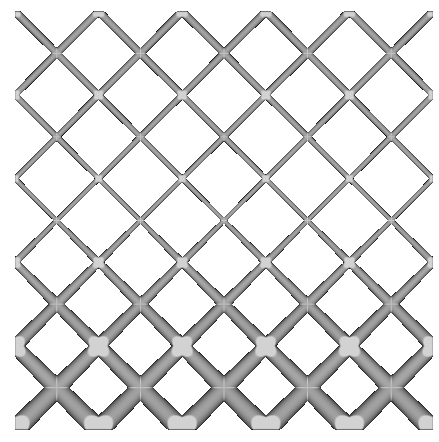}}
    \end{tabular}
    \caption{\hl{Comparative images of a homogeneous (top row) and heterogeneous (bottom row) strut-based body centered cubic lattice structure.}}
    \label{fig:BCC_homo_hetero}
\end{figure}

\begin{figure}[htbp]
    \centering
    \captionsetup{justification=centering}
    \begin{tabular}{@{}lcc@{}}
        & \textbf{Periodic} & \textbf{Pseudo-periodic} \\
         \rotatebox[origin=c]{90}{\textbf{Homogeneous}} & \raisebox{-.5\height}{\includegraphics[scale=0.5]{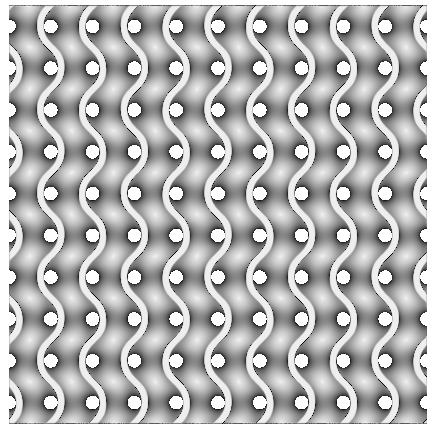}}&  \raisebox{-.5\height}{\includegraphics[scale=0.42]{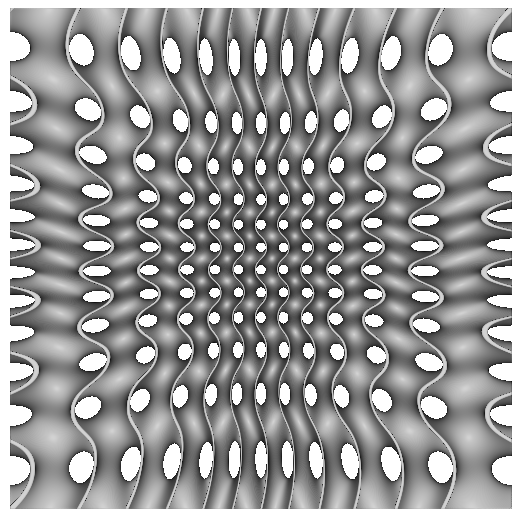}} \\
         \rotatebox[origin=c]{90}{\textbf{Heterogeneous}} & \raisebox{-.5\height}{\includegraphics[scale=0.5]{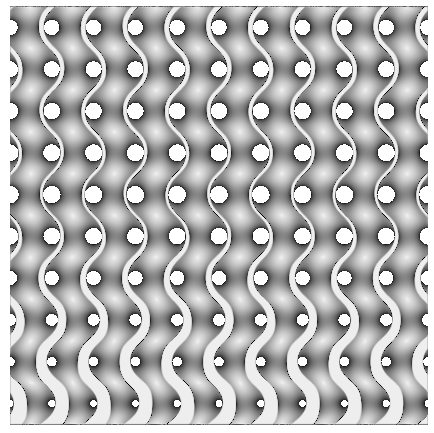}}&  \raisebox{-.5\height}{\includegraphics[scale=0.42]{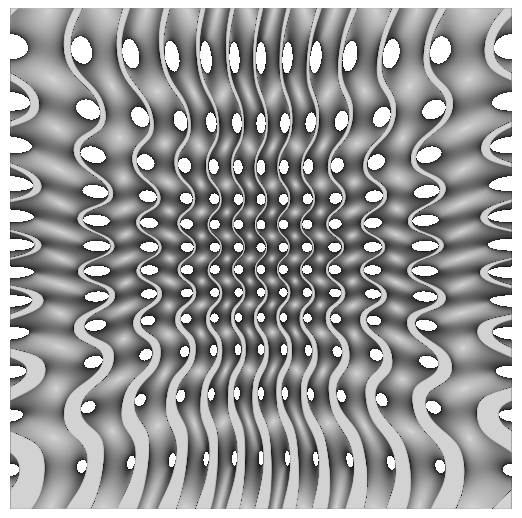}}
    \end{tabular}
    \caption{\hl{Comparative cross-sectional images of various TPMS Schoen gyroid lattice structures classified as periodic, pseudo-periodic, homogeneous, and heterogeneous.}}
    \label{fig:P_PP_homo_hetero}
\end{figure}

Cellular lattices can then be further categorized into periodic and pseudo-periodic lattices that are of homogeneous or heterogeneous organizations (Figure~\ref{fig:lattice_types}) \cite{dong2017survey}. The periodicity of a lattice refers to the size of the unit-cells throughout the structure and homogeneity refers to the thickness of the unit-cell elements such as struts and walls. Therefore, a periodic lattice would have a uniform unit-cell size throughout its structure and a pseudo-periodic structure would have variable unit cell size, as shown in Figure~\ref{fig:TPMS_diamond} \hl{for a triply period minimal surface (TPMS) Schwarz diamond (D) lattice structure. Both periodic and pseudo-periodic lattices can be further categorized into homogeneous or heterogeneous lattices. Homogeneous lattice structures have a uniform strut and/or wall thickness and heterogeneous lattice structures have gradients in the strut and/or wall thickness, as shown in Figure}~\ref{fig:BCC_homo_hetero} \hl{for a strut-based body centered cubic lattice structure. A visual comparison between periodic and pseudo-periodic lattice structures with a homogeneous and heterogeneous organization is given in Figure}~\ref{fig:P_PP_homo_hetero} \hl{for TPMS Schoen gyroid lattices}.

\subsubsection*{\hl{Organization driven classification}}

Another form of categorization for lattice structures is by their cellular organization, as shown in Figure~\ref{fig:lattice_types}. The four commonly known cellular organizations of lattice structures are strut-based, surface-based, plate-based, and honeycomb lattices \cite{zheng2016multiscale, al2019multifunctional, tancogne20183d, bauer2017nanolattices, wadley2006multifunctional}, as shown in Figure~\ref{fig:lattice_types}. Strut-based lattices are generated by determining unit cell size, the number nodes located throughout the unit cell, and the number and configuration of connectors linking nodes to each other. Porosity of the lattice may be controlled directly, or strut thickness may be selected as the control variable for lattice density. Surface-based lattices consist of a locus of points defined by a function. The most commonly used and discussed surface lattice family are triply period minimal surface (TPMS) lattices. TPMS lattices are defined by implicit functions for which the function has a constant value. \hl{Plate-based lattices can be derived either from strut-based lattices by placing plates between two adjacent trusses} \cite{liu2021mechanical}, \hl{or by placing plates on the closest packed planes of lattices derived from crystal structures} \cite{berger2017mechanical, tancogne20183d, crook2020plate}. \hl{Plate-based lattices are known to have significantly higher stiffness when compared to truss-based lattices} \cite{berger2017mechanical, tancogne2019high, andrew2021impact}. \hl{Honeycomb lattices consist of plates that form the edges of unit cells. The unit cells for honeycomb lattices can have either a triangular, hexagonal} (Figure~\ref{fig:lattice_types}), \hl{square, or related shapes forming the 2D cross-section. These unit cells are then repeated in a 2D space to obtain a cellular solid, and then extruded in a 3D space to obtain a lattice structure, such as the hexagonal honeycomb structure shown in} Figure~\ref{fig:lattice_types} \cite{wadley2006multifunctional}. \hl{All types of lattice structures may be periodic or pseudo-periodic and heterogeneous or homogeneous.} 

\subsubsection{Stochastic lattices}
Stochastic, or random, lattice structures consist of irregular and non-periodic cells resulting in a network of interconnected struts, surfaces, and/or plates. Unlike other lattice structures there are no distinct cellular features that are repeated in 3D space and each cell contains a unique configuration of struts and nodes. Stochastic lattices have superior performance under both compressive and shear loading when compared to regular lattice structures \cite{maliaris2017mechanical}. However, due to their complex design, there are not as many readily available tools for the design and implementation and therefore they are not as commonly examined in literature. The final lattice type to be examined are spinodoid lattices \cite{kumar2020inverse}, which are a subset of stochastic lattices and surface lattices. The major differentiating characteristic of spinodal lattices is that they are non-periodic. This allows for a larger design space and more achievable control of the directional mechanical properties. The benefit of spinodal surfaces is that they are immune to the symmetry-breaking defects present in cellular lattices, thus improving their mechanical properties \cite{kumar2020inverse}. 

\subsubsection{Mechanical properties of PBF lattice structures}

\begin{figure}[htbp]
    \centering
    \captionsetup{justification=centering}
    \includegraphics[width=16cm,keepaspectratio]{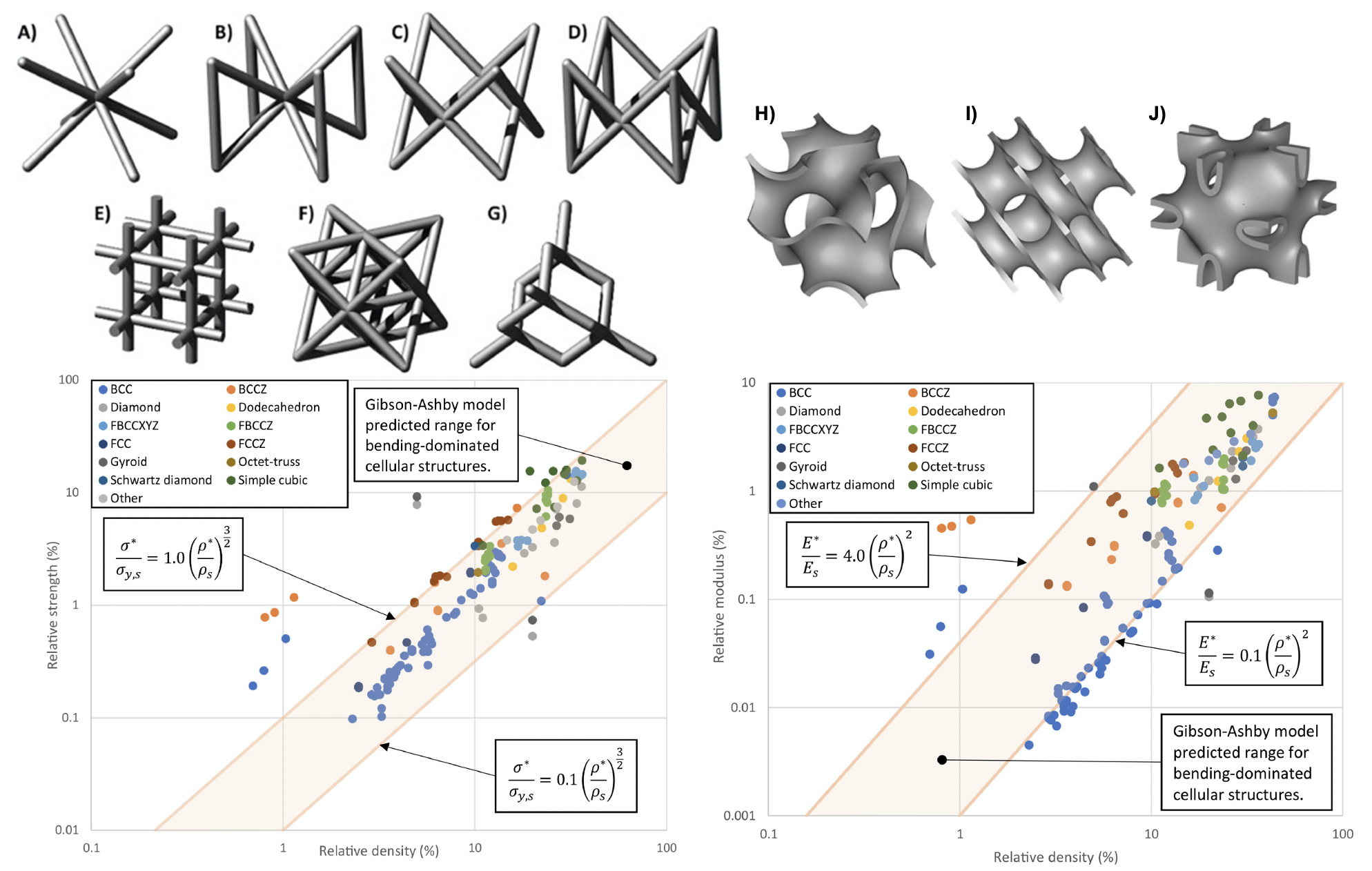}
    \caption{\hl{Experimental data comparing the compressive strength (left) and Young's modulus (right) for common cellular lattice structures to those predicted by the Gibson-Ashby model. Cellular lattice structures examined include: BCC (A), BCCZ (B), FCC (C), FCCZ (D), cubic (E), Octet-truss (F), diamond (G), Schoen gyroid (H), Schwarz diamond (I), and Neovius (J). Original figures obtained from} \cite{maconachie2019slm}.}
    \label{fig:gibson_ashby_model}
\end{figure}

\hl{Mechanical properties of lattice structures, such as compressive strength and elastic modulus, typically refer to the macroscopic properties of a unit cell or group of units cells which differs from the mechanical properties of the bulk modulus for a given material} \cite{zadpoor2019mechanical}. \hl{Typically, the relationship between the mechanical properties of lattice structures and those of the bulk modulus can be described using a ratio which relates apparent density of a cellular lattice to the density of the structure's material. Therefore, as apparent density of the lattice structure decreases, or designed lattice porosity increases, mechanical properties of the lattice structure decrease. The Gibson-Ashby model is the most widely accepted model used for predicting the mechanical properties of cellular-lattice structures }\cite{gibson2003cellular}. \hl{Maconachie et al. conducted a detailed review of mechanical properties of lattice structures produced by LPBF} \cite{maconachie2019slm}. \hl{They used the Gibson-Ashby model to predict mechanical properties of common cellular lattice structures and compared the findings to experimental data} (Figure~\ref{fig:gibson_ashby_model}). \hl{They found that the Gibson-Ashby model was useful in predicting mechanical properties of cellular lattice structures produced trough LPBF.} 

\section{Review methodology}
A comprehensive literature review was completed to better understand how lattice parameters are controlled in additively manufactured titanium and titanium alloy parts aimed at replacing or augmenting bone. In order to collect the most relevant data, all powder bed additive manufacturing processes were considered, pure titanium and titanium alloys were considered, and all study types were considered; however, studies were only included when bone was the target tissue for replacement, repair and/or augmentation, to enhance the relevant scope of the designed architectures. A total of 50 journal articles fit the above criteria and the effect of lattice design parameters on mechanical properties was extracted and examined \cite{mobbs2017utility, kim2017sacral, choy2017reconstruction, xu2016reconstruction, taniguchi2016effect, de2013bone, hilton2017additive, schouman2016influence, arabnejad2017fully, wu2013porous, biemond2013bone, xue2007processing, van2013selective, srivas2017osseointegration, wieding2015biomechanical, van2012effect, otsuki2006pore, ghouse2019design, fousova2017promising, arabnejad2016high, moiduddin2017structural, harrysson2008direct, wong2015one, wang2017mapping, lin2013additive, barbas2012development, wieding2014numerical, du2018ti6al4v, arjunan2020mechanical, soro2019investigation, alabort2019design, zhang2018biomimetic, bartolomeu2021selective, balci2021reproducibility, xiong2020rationally,dallago2021role, bari2019extra, heinl2008cellular, liu2018mechanical, yan2015ti, el2020design, wang2020electron, zhang2020biomechanical, phan2016application, taheri2016achieving, murr2011microstructure, ge2020microstructural}.

Existing literature describing additive manufactured titanium implants for bone replacement fits into two main categories of critical design focus: studies focused on improving osseointegration and studies focused on targeted mechanical properties. Osseointegration refers to bone’s ability to grow on the surface of the implants and infiltrate the porous implant to improve implant fixation. In general, literature focused on osseointegration was found to have fewer reported AM and lattice parameters provided and often focused on \textit{in vivo} results in animals or human case studies. The primary target audience for this category seems to be medical and academic researchers interested in bone tissue mechanics, growth and healing and secondarily, the additive manufacturing community. The other category of literature aims at matching the mechanical properties of bone by controlling the lattice design parameters and by controlling the printing process parameters of the respective technologies. Literature focused on matching mechanical properties of lattice structures to bone, primarily targets the additive manufacturing community with implications for bone tissue and device design being secondary suggestions. It is noteworthy to highlight this lack of apparent synergy between the two categories; such synergy is required to ensure advancements in this field. 

A wide cross-section of lattice design information was collected from the bone and AM focused journals and was collated in the associated Data in Brief. Key parameters collected include: Young's modulus, compressive strength, lattice porosity, pore size, feature thickness, lattice type and material used. When studies compared more than one lattice design parameter, all relevant data points were collected in order to make the most robust comparison possible. Data points were then plotted on alongside Ashby plots for trabecular and cortical bone to assist in making recommendations for future lattice designs focused on titanium implant designs for bone repair and replacement (CES EduPack software, Granta Design Limited, Cambridge, UK, 2009).

\section{Results and discussion}
\subsection{Lattice considerations}

\subsubsection{Porosity}
\begin{figure}[htbp]
    \centering
    \captionsetup{justification=centering}
    \includegraphics[width=15cm,keepaspectratio]{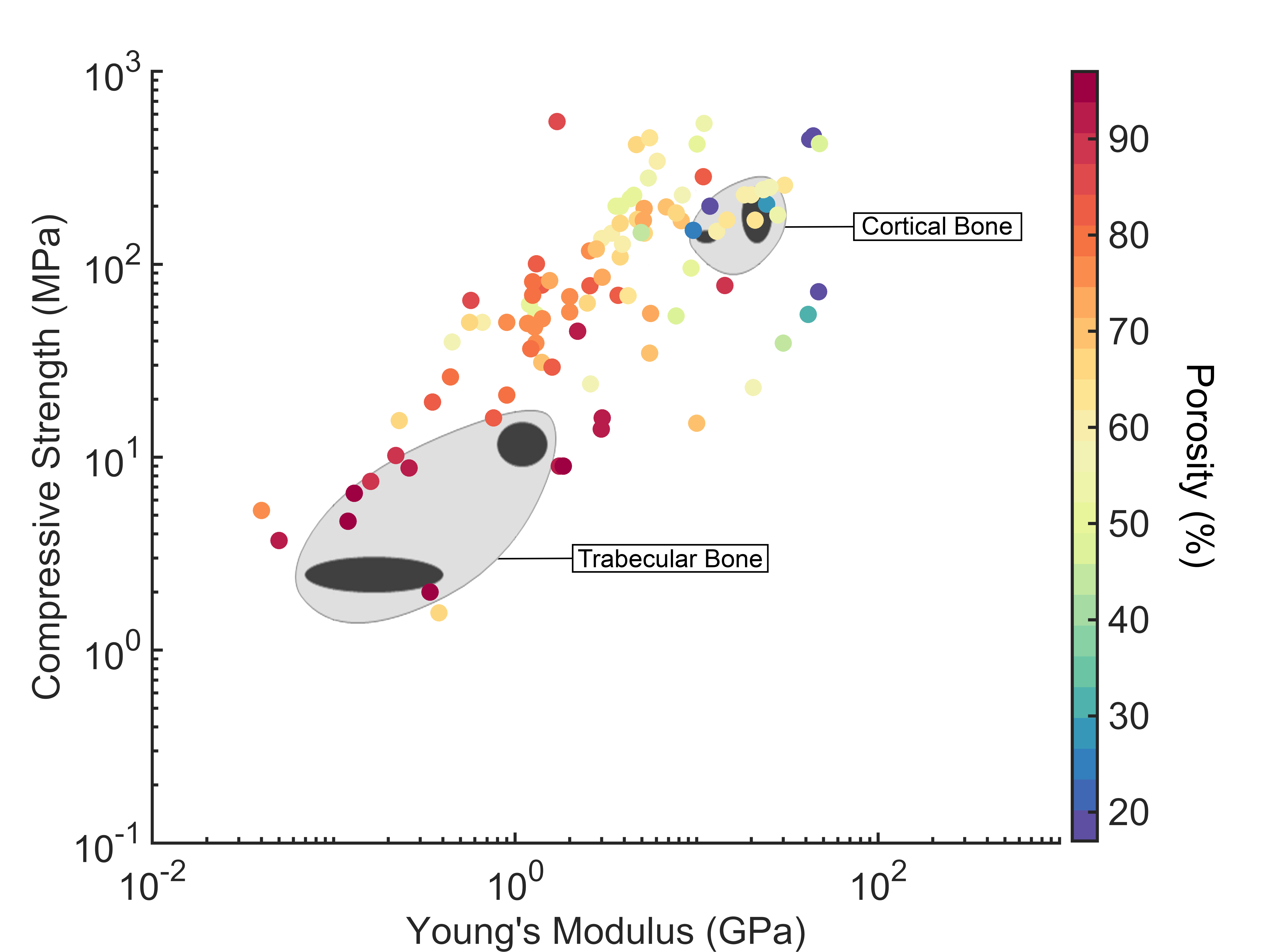}
    \caption{Porosity is the most common lattice parameter reported in literature. Compressive strength and Young's modulus of additively manufactured lattice structures were plotted over the Ashby plots of human trabecular and cortical bone. Lattice porosity ranged from 15-97\% and was plotted in a gradient to depict how best to design for material property matching \cite{taniguchi2016effect,de2013bone, wu2013porous, xue2007processing, van2013selective, srivas2017osseointegration, wieding2015biomechanical, ghouse2019design, zhao2018effect, fousova2017promising, arabnejad2016high, moiduddin2017structural, taheri2016achieving, marin2013characterization, barbas2012development, du2018ti6al4v, arjunan2020mechanical, alabort2019design, zhang2018biomimetic, xiong2020rationally, dallago2021role, heinl2008cellular, liu2018mechanical, yan2015ti, ge2020microstructural, el2020design, wang2020electron}.}
    \label{fig:porosity}
\end{figure}

The exercise of understanding trabecular and cortical bone porosity provides insight into the porosity required to match the structural properties of bone through titanium and Ti-6Al-4V latticing \cite{taniguchi2016effect,de2013bone, wu2013porous, xue2007processing, van2013selective, srivas2017osseointegration, wieding2015biomechanical, ghouse2019design, zhao2018effect, fousova2017promising, arabnejad2016high, moiduddin2017structural, taheri2016achieving, marin2013characterization, barbas2012development, du2018ti6al4v, arjunan2020mechanical, alabort2019design, zhang2018biomimetic, xiong2020rationally, dallago2021role, heinl2008cellular, liu2018mechanical, yan2015ti, ge2020microstructural, el2020design, wang2020electron}. The most common lattice parameter reported in literature focused on additively manufactured titanium and titanium-alloy lattice structures was macro-scale porosity or void fraction. Designed lattice porosity varied from 15\% to 97\% with the majority of studies reporting a designed porosity between 50-70\%. Titanium and titanium alloy lattices within this designed porosity range were successful in matching the stiffness of cortical bone which is known to be 5-15\% porous. However, few were successful in matching mechanical properties of trabecular bone, as depicted in Figure \ref{fig:porosity}. Trabecular bone is ~70-90\% porous, and through this review, it was determined that titanium and Ti-6Al-4V lattices must have a designed porosity of >80\% to replicate mechanical properties of trabecular bone. These findings indicate that matching material properties of titanium and Ti-6Al-4V via latticing may be challenging with existing AM technologies.

\subsubsection{Pore size}
\begin{figure}[htbp]
    \centering
    \captionsetup{justification=centering}
    \includegraphics[width=15cm,keepaspectratio]{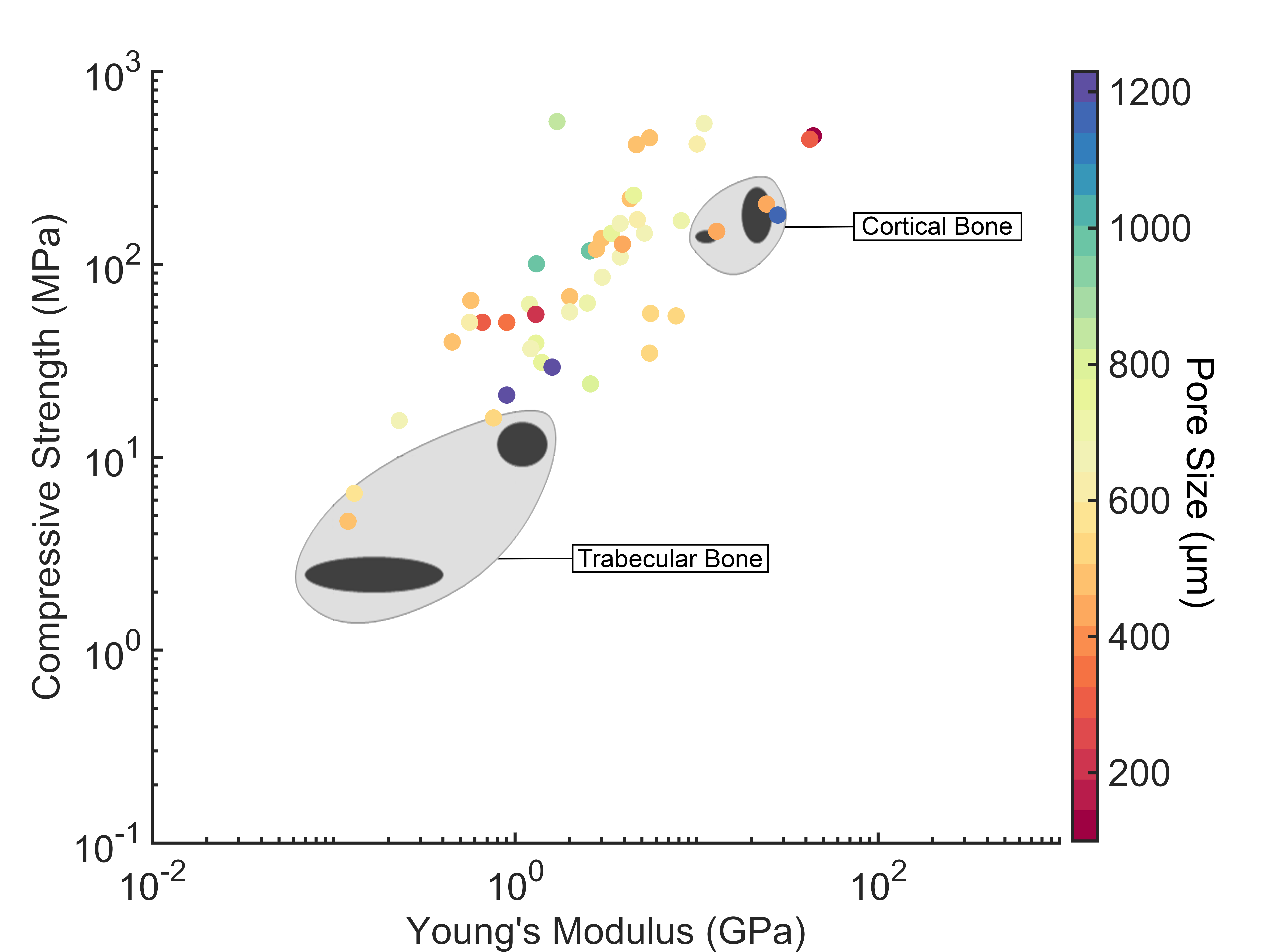}
    \caption{Designed pore size of additively manufactured Ti and Ti-6Al-4V lattice structures was plotted in a gradient over a compressive strength versus Young's modulus Ashby plot for human trabecular and cortical bone tissues to depict how best to design for material property matching \cite{taniguchi2016effect, de2013bone, wu2013porous, xue2007processing, srivas2017osseointegration, wieding2015biomechanical, ghouse2019design, zhao2018effect, arabnejad2016high, moiduddin2017structural, marin2013characterization, barbas2012development, alabort2019design, zhang2010osteogenic, xiong2020rationally, heinl2008cellular,yan2015ti, ge2020microstructural}.}
    \label{fig:poresize}
\end{figure}

Two-dimensional micro-architecture measurements of bone can also be described in terms of lattice parameters. When considering TbSp as a surrogate bone pore size, Hildebrand et al. reported a range from 638 \textmu m in the femoral head to 854 \textmu m in the lumbar spine \cite{hildebrand1999direct}. In the literature focused on osseointegration for additively manufactured titanium lattice designs, the lattice parameter most commonly reported was pore size. The pore sizes reported ranged from 100-1500 \textmu m. Recommendations for tailoring pore size to optimize bone in-growth or osseointegration were consistent and conclusions were drawn surrounding an acceptable range for optimal boney ingrowth. A minimum pore size of 200 \textmu m should be considered to allow for initial cell adhesion \cite{xue2007processing}. However, to maximize cell proliferation and limit cell occlusion, large pore sizes >1000 \textmu m are recommended \cite{zhao2018effect}. Therefore, a functionally graded lattice which combines small pores for initial cell attachment and large pores to avoid cell occlusion would account for both recommendations \cite{van2012effect,yadroitsava2019bone}. Pore size did not exhibit a trend with respect to compressive strength and Young’s modulus as seen in Figure \ref{fig:poresize} \cite{taniguchi2016effect, de2013bone, wu2013porous, xue2007processing, srivas2017osseointegration, wieding2015biomechanical, ghouse2019design, zhao2018effect, arabnejad2016high, moiduddin2017structural, marin2013characterization, barbas2012development, alabort2019design, zhang2010osteogenic, xiong2020rationally, heinl2008cellular,yan2015ti, ge2020microstructural}. Therefore, pore size should be viewed as a design parameter for biological reaction rather than mechanical function.

\subsubsection{Feature Thickness}
\begin{figure}[htbp]
    \centering
    \captionsetup{justification=centering}
    \includegraphics[width=15cm,keepaspectratio]{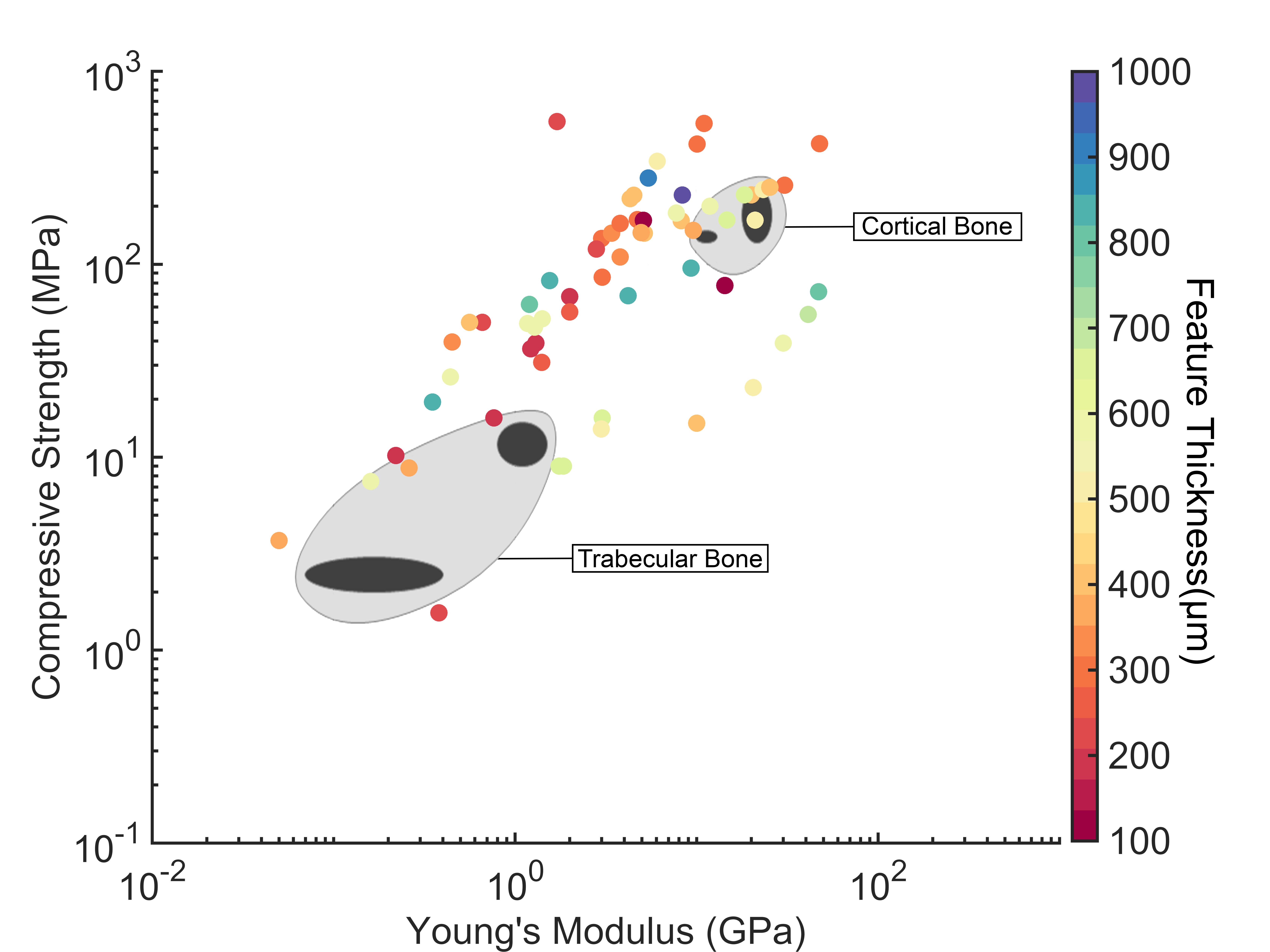}
    \caption{Feature thickness was plotted in a gradient over a compressive strength versus Young's modulus Ashby plot for human trabecular and cortical bone tissues to depict how best to design for material property matching \cite{taniguchi2016effect, de2013bone, van2013selective, srivas2017osseointegration,wieding2015biomechanical, ghouse2019design, fousova2017promising, arabnejad2016high, moiduddin2017structural, taheri2016achieving, du2018ti6al4v, zhang2018biomimetic, xiong2020rationally, dallago2021role, el2020design, wang2020electron}.}
    \label{fig:featuresize}
\end{figure}

Trabecular thickness can be related to feature, strut or wall thickness. Across the human skeleton, trabecular thickness varies from roughly 120-200 \textmu m \cite{hildebrand1999direct}. This is lower than 400 \textmu m, or the minimum feature thicknesses typically recommended for powder bed fusion. This is most likely due to the part resolution that can be obtained through current additive manufacturing technologies. While there were no strong trends in the effect of feature thickness on compressive strength and Young’s modulus for the feature size range captured in these studies, decreasing feature thickness is one way to control lattice porosity, which is critical to manipulating mechanical properties, as seen in Figure \ref{fig:featuresize} \cite{taniguchi2016effect, de2013bone, van2013selective, srivas2017osseointegration,wieding2015biomechanical, ghouse2019design, fousova2017promising, arabnejad2016high, moiduddin2017structural, taheri2016achieving, du2018ti6al4v, zhang2018biomimetic, xiong2020rationally, dallago2021role, el2020design, wang2020electron}. 

\subsubsection{Lattice type}
\begin{figure}[htbp]
    \centering
    \captionsetup{justification=centering}
    \includegraphics[width=15cm,keepaspectratio]{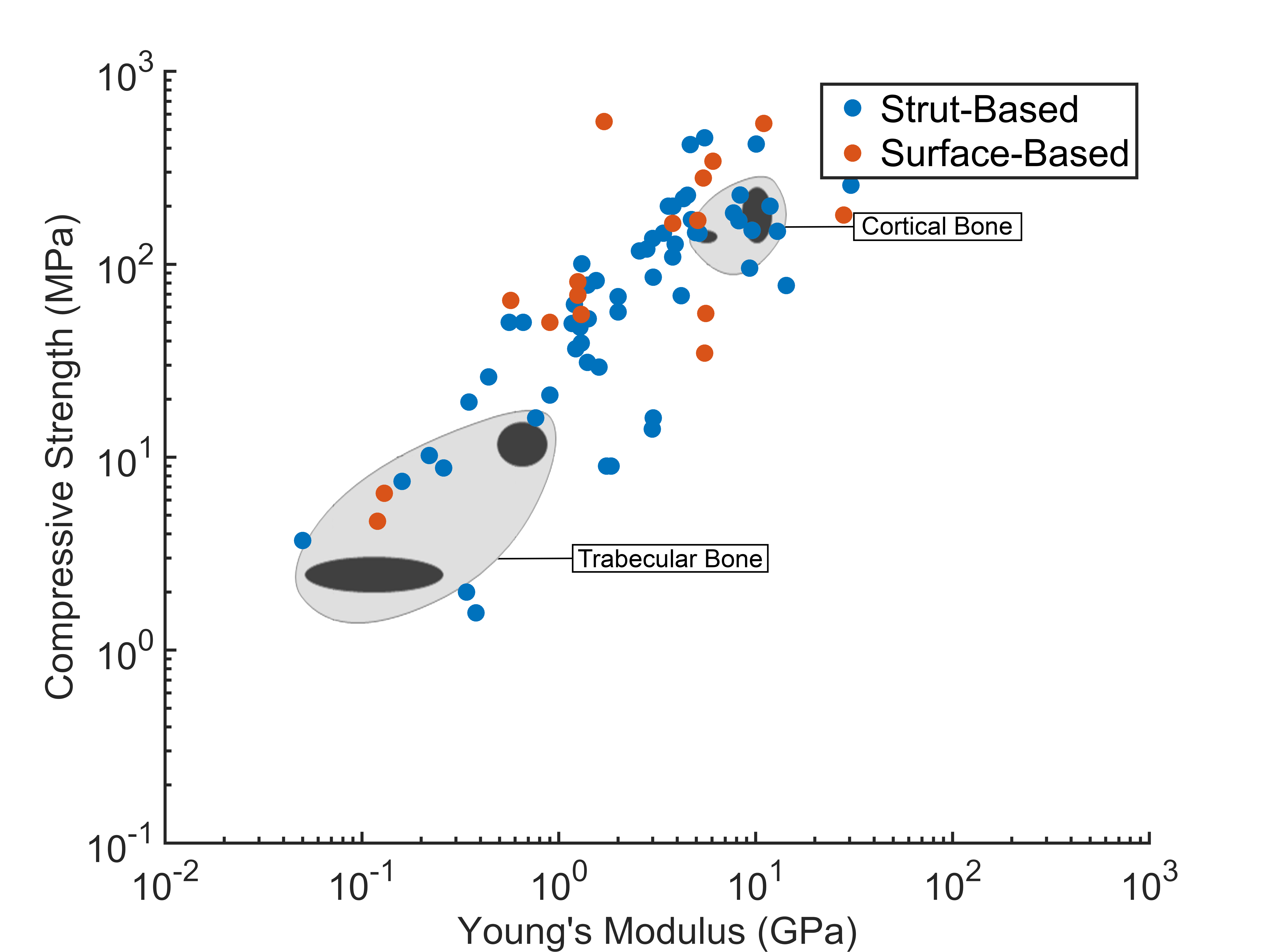}
    \caption{Lattice type, strut- vs surface-based, was plotted in a gradient over a compressive strength versus Young's modulus Ashby plot for human trabecular and cortical bone tissues to depict how best to design for material property matching \cite{taniguchi2016effect, de2013bone, van2013selective, ghouse2019design, zhao2018effect, fousova2017promising, arabnejad2016high, moiduddin2017structural, barbas2012development, du2018ti6al4v, alabort2019design, zhang2018biomimetic, xiong2020rationally, dallago2021role, heinl2008cellular, liu2017elastic, yan2015ti, ge2020microstructural, el2020design}.}
    \label{fig:strutsurface}
\end{figure}

When compared to strut-based lattices, surface-based lattices, such as TPMS structures, allow for better osseointegration \cite{yadroitsava2019bone}. This is thought to be due to the increased surface area available in surface-based lattices for cellular adhesion. However, the designed lattice porosity needed to match the mechanical properties of trabecular bone requires very thin wall thickness. Surface lattices also have lower stress concentrations under angular load simulation, which may make them further suitable for bone implants \cite{yadroitsava2019bone}. A recent study from Alabort et al. showed promising results for reaching the mechanical properties of trabecular bone through the use of TPMS surface lattices, specifically Schwartz’s diamond surface structures \cite{alabort2019design}. In this review, lattice type had no noticeably influence on compressive strength nor Young’s modulus of titanium and Ti-6Al-4V lattice structures aimed at human bone replacement, see Figure \ref{fig:strutsurface} \cite{taniguchi2016effect, de2013bone, van2013selective, ghouse2019design, zhao2018effect, fousova2017promising, arabnejad2016high, moiduddin2017structural, barbas2012development, du2018ti6al4v, alabort2019design, zhang2018biomimetic, xiong2020rationally, dallago2021role, heinl2008cellular, liu2017elastic, yan2015ti, ge2020microstructural, el2020design}. 

\subsubsection{Material selection}
\begin{figure}[htbp]
    \centering
    \captionsetup{justification=centering}
    \includegraphics[width=15cm,keepaspectratio]{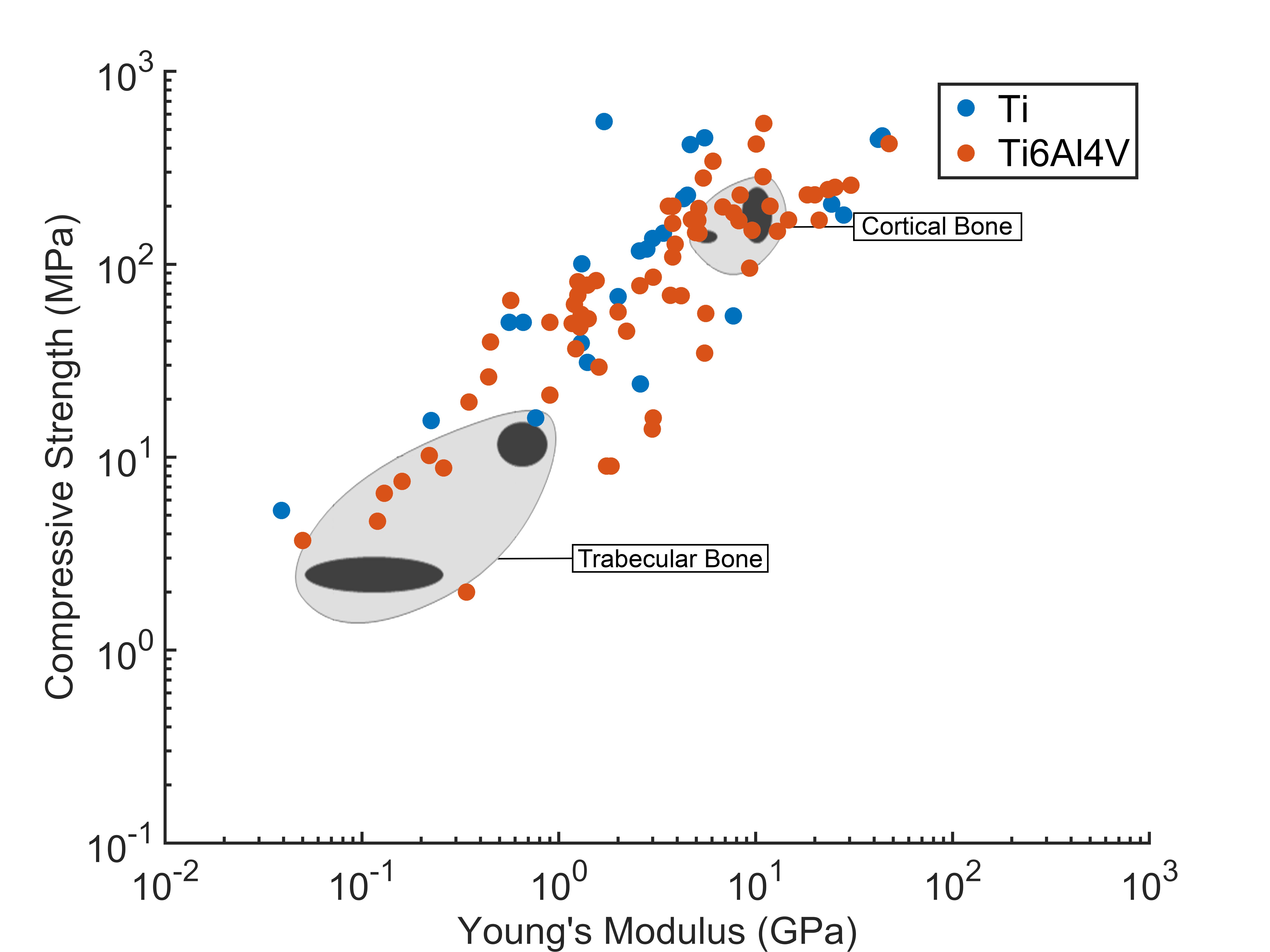}
    \caption{Material type, Ti vs Ti-6Al-4V, was plotted in a gradient over a compressive strength versus Young's modulus Ashby plot for human trabecular and cortical bone tissues to depict how best to design for material property matching \cite{taniguchi2016effect,de2013bone, wu2013porous, xue2007processing, srivas2017osseointegration, wieding2015biomechanical, ghouse2019design, zhao2018effect, fousova2017promising, arabnejad2016high, moiduddin2017structural, marin2013characterization, barbas2012development, du2018ti6al4v, arjunan2020mechanical, alabort2019design, zhang2018biomimetic, xiong2020rationally, dallago2021role, heinl2008cellular, liu2018mechanical, yan2015ti, ge2020microstructural, el2020design, wang2020electron}.}
    \label{fig:material}
\end{figure}

Finally, material choice was examined and Ti and Ti-6Al-4V lattices were compared for their ability to achieve comparable Young's Modulus and compressive strength to human bone tissues \cite{taniguchi2016effect,de2013bone, wu2013porous, xue2007processing, srivas2017osseointegration, wieding2015biomechanical, ghouse2019design, zhao2018effect, fousova2017promising, arabnejad2016high, moiduddin2017structural, marin2013characterization, barbas2012development, du2018ti6al4v, arjunan2020mechanical, alabort2019design, zhang2018biomimetic, xiong2020rationally, dallago2021role, heinl2008cellular, liu2018mechanical, yan2015ti, ge2020microstructural, el2020design, wang2020electron}. Despite having slightly different bulk modulii, Ti and Ti-6Al-4V lattices did not differ in ability to reach bone properties, see Figure \ref{fig:material}. This may be due to other lattice design decisions, such as porosity, pore size and feature thickness, that were made to tailor overall mechanical properties.

\subsubsection{Considerations for fatigue strength}

\hl{Fatigue strength is another mechanical property that is important to characterize for bone replacement and repair applications due to the cyclic nature of \textit{in vivo} implant loading. Fatigue strength of additively manufactured lattice structures have been examined with manufacturing defects, bulk material modulus, designed lattice porosity, and lattice geometry all playing key roles} \cite{zargarian2016numerical}. \hl{To date, there has been limited investigations into fatigue strength of titanium lattice structures for bone replacement and repair specifically. Dallago et al. examined the effects of unit-cell size and strut orientation on fatigue strength of Ti-6Al-4V lattice structures produced through laser powder bed fusion }\cite{dallago2021role}. \hl{They found that fatigue strength of lattice structures can be improved for bone replacement by optimizing build direction and junction geometry. They suggest that to improve fatigue strength load bearing struts should be printed perpendicular to the build plate and unit cell size should be up-scaled to decrease manufacturing defects associated with reduced fatigue strength }\cite{dallago2021role}. \hl{These recommendations should be taken into account when designing for titanium lattices aimed at human bone replacement and repair, particularly when the fatigue life requirements for certain implants are to withstand loads without failure for over 5 million cycles, as noted in ISO 7206-4 for hip implants} \cite{isoimplants}.

\subsection{Manufacturability considerations}
The potential of powder bed fusion AM technologies such as LPBF and EB-PBF to manufacture parts with higher geometric complexity compared to traditional manufacturing, makes them well suited for fabricating lattice structures mimicking bone properties. The complex features involved in the design of most lattice structures tests the manufacturability limits of LPBF and EB-PBF. This is mainly because most lattice structures used for bone replacements require fine feature sizes, particularly to replace trabecular bone. The minimum feature size strongly depends upon the beam spot size used which is generally between 50-100 \textmu m for LPBF and >200 \textmu m for EB-PBF. Additionally, lattice structure designs generally incorporate numerous overhanging features within a unit cell, which are challenging to produce by both LPBF \cite{charles2020dimensional, druzgalski2020process} and EB-PBF \cite{wang2018effect, newton2020feature, cooper2018contact}. \hl{For LPBF in particular, surfaces with an overhang angle of} $<$\ang{30} \hl{(with respect to the PBF build plate) are considered non self-supporting leading to print failures or significant distortion in the manufactured lattice structures} \cite{maconachie2019slm}. Pushing the design boundaries in LPBF and EB-PBF to achieve lattice architectures tailored for bone replacement and augmentation necessitates an understanding of the three main categories of manufacturability challenges which arise in these AM technologies - defects (micro-porosity within the manufactured lattice structure), surface roughness, and geometric fidelity. 

\subsubsection{Porous defects}
Due to the fatigue strength and stiffness requirements associated with manufacturing titanium-based bone replacements, understanding defects is important, as they directly impact both stiffness and fatigue life of a given AM part. It is well documented in AM literature that defects are particularly deleterious for fatigue properties \cite{soro2021quasi, du2020effects, echeta2020review}. In most of the articles reviewed in this work, stiffness values and lattice design details were commonly reported for titanium alloy lattice structures used for bone replacements, but studies into the defects within the lattice structures and their effects on fatigue life were less frequently reported.

\begin{figure}[htbp]
    \centering
    \captionsetup{justification=centering}
    \includegraphics[width=15cm,keepaspectratio]{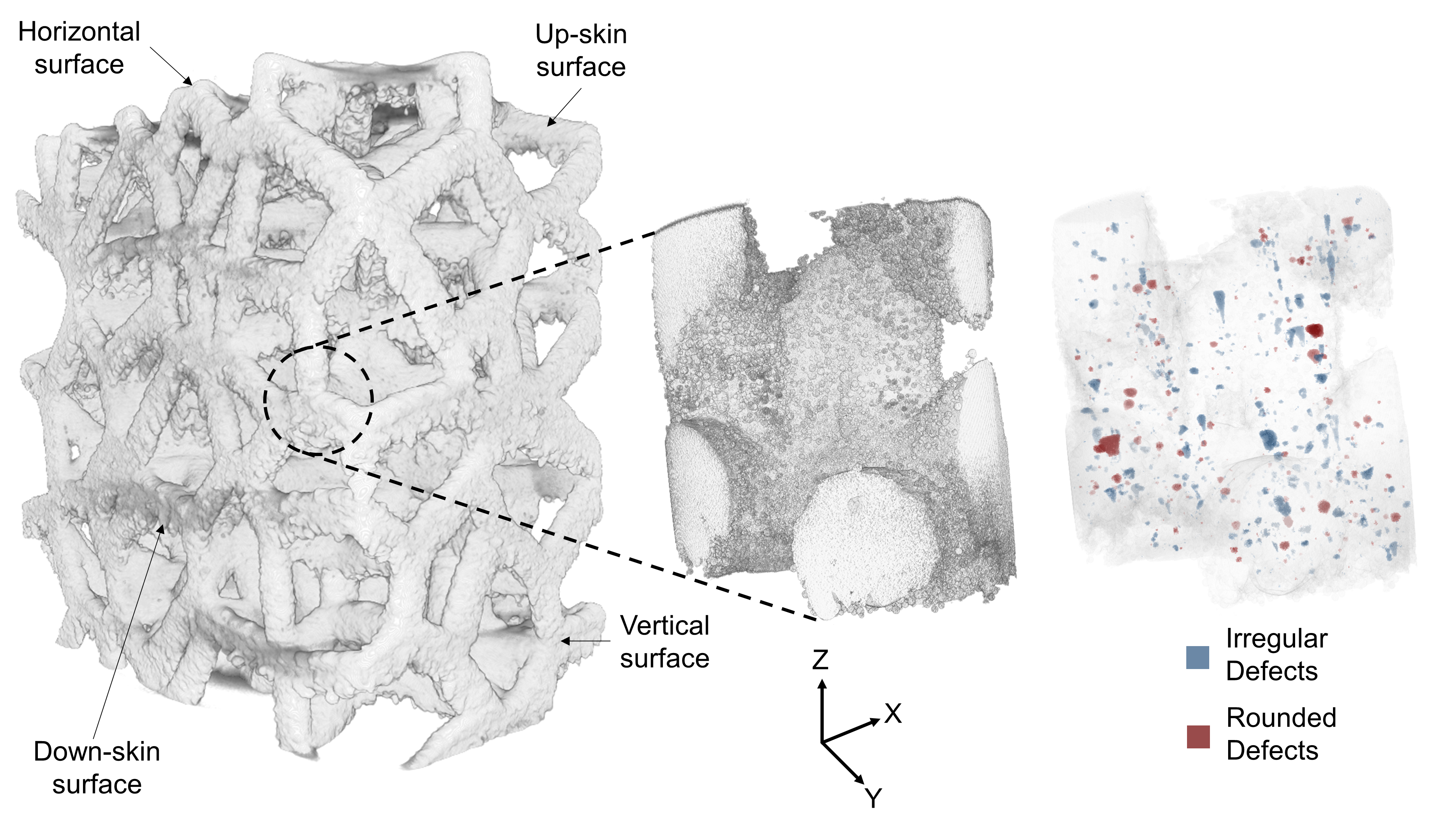}
    \caption{A three-dimensional XCT visualization of the different types of surfaces observed in a laser powder bed fusion Ti-6Al-4V Voronoi lattice structure with respect to the build orientation along the Z-axis (left), a high resolution XCT image of a portion of the Voronoi lattice structure (center), visualization of the defects inside the printed lattice structure (right).}
    \label{fig:XCT}
\end{figure}

The amount of defects and their typical morphology observed in powder bed fusion (PBF) AM depends on the process parameters used for manufacturing a given lattice structure. A low energy input is typically associated with the formation of irregularly shaped lack of fusion defects \cite{patel2020melting}, with a high aspect ratio (width/depth) which are are known to be more detrimental to the fatigue strength of lattice structures \cite{du2020effects}, when compared to rounded keyhole defects typically associated with high energy inputs \cite{patel2020melting}. It is important to note that the presence of rounded or irregularly-shaped defects is not only dependent upon the energy input, but rather requires an understanding of PBF process parameters, particularly power, beam velocity, beam spot size, powder layer thickness, and hatching distance. It is quite possible to obtain irregularly shaped, lack of fusion defects within a lattice structure that uses high energy keyhole mode parameters, as shown by an X-ray computed tomography visualization of the defect space within a Ti-6Al-4V Voronoi lattice structure manufactured by LPBF in Figure \ref{fig:XCT}. A summary of the LPBF processing details and XCT measurements of the lattice structure are provided in the associated Data in Brief.

\subsubsection{Surface roughness} \label{roughness}
Surfaces in PBF parts are generally identified with their orientation with respect to the build plate used for manufacturing as shown by the left image in Figure \ref{fig:XCT}. The four type of surfaces shown in Figure \ref{fig:XCT} include - horizontal up-facing surfaces that are parallel to the build plate (along the XY plane), vertical surfaces that are perpendicular to build plate (along the Z axis and also known as side-skin surfaces), upward facing surfaces (known as up-skin surfaces) which are typically on an incline, but facing upwards, and downward facing surfaces (known as downskin surfaces) \cite{patel2020towards}. Since most lattices used for bone replacement are comprised of a combination of all four surface types, these surfaces are distinctly different contributors towards the final surface roughness in the printed lattice structures. Side-skin (vertical) surfaces \cite{patel2020towards, abele2015analysis} and down-skin surfaces \cite{chen2018surface, tian2017influences, echeta2020review} are associated with a higher number of challenges when trying to obtain lower surface roughness values in as-printed PBF lattices, when compared to horizontal up-facing and up-skin surfaces.

In PBF AM processes, an interplay between process parameters, build file characteristics, machine characteristics, and powder characteristics drive the final surface topography of a given lattice structure. More precisely, processing parameters such as power, scan speed and layer thickness, build file characteristics such as feature geometry, feature orientation, feature location on the build plate, and beam path strategy, machine and energy source characteristics such as beam spot size, laser beam quality, and gas flow (for LPBF), and powder morphology and size distribution are some of the primary drivers for roughness of a given lattice structure.

Out of the four influencing factors, PBF process parameters are known to have the greatest effect on roughness and are also the most readily controllable for a given lattice. Horizontal and up-skin surface roughness possibly depend upon the overall size of the melt pools \cite{snyder2020understanding, tian2019experimental}, where the roughness on these surface features generally includes visibility of the melt pool tracks alongside partially fused adhered powder, particularly for the up-skin surfaces. Down-skin surface features include partially fused adhered powder as sources of coarse roughness and dross (at higher energy inputs); overall, the down-skin roughness strongly depends on the melt pool depth \cite{tian2017influences, snyder2020understanding}. A melt pool depth close to the powder layer thickness is generally considered to be useful for lowering down-skin surface roughness values \cite{tian2017influences}. Remelting scans of horizontal, up-skin, and down-skin surfaces are known to improve the roughness values of these surfaces \cite{snyder2020understanding}. Side-skin (vertical) surfaces are an exception, wherein remelting would generally not help further improve surface roughness, when compared to a well-executed first side-skin scan \cite{calignano2018investigation}. Side-skin surfaces are generally dominated by partially fused adhered powder in PBF, but the effects of powder can be reduced by an appropriate energy input selection which enables a dominance of melt pool track features on the side-skin that are associated with lower surface roughness values \cite{patel2020melting, abele2015analysis}.

Surface roughness of AM titanium lattices for bone replacement applications has been previously examined in literature. Webster and Ejiofor examined osteoblast proliferation on 90-95\% dense Ti and Ti-6Al-4V structures. They reported that osteoblasts prefer titanium surfaces with nanometer topology features \cite{webster2004increased}. Conversely, Zhang et al. reported that a surface roughness of 1-2 \textmu m improved osteogenic properties of titanium bone implants \cite{zhang2010osteogenic}. There is still no consensus on optimal surface topology and roughness for osseointegration and adhesion, as changes in the microstructure and nanostructure of lattice surfaces both influence the cell response in bone tissues \cite{gui2018effect}.

\subsubsection{Geometric fidelity of lattice features}
The fine and complex features involved in the CAD of most lattices used for bone replacements test the manufacturability limits of both LPBF and EB-PBF. This implies that a lattice structure with a fine feature size might print successfully, but not conform to the original CAD, leading to dimensional inaccuracies. Inaccuracies in additively manufactured lattices would directly impact their performance as bone replacements, since the reality of the AM structure would differ from the modelling efforts and design decisions used to determine the feature size and morphology of the lattice. Attempts at reporting the geometric fidelity of AM lattices, however, have been scarce.

\begin{figure}[htbp]
    \centering
    \captionsetup{justification=centering}
    \includegraphics[width=12cm,keepaspectratio]{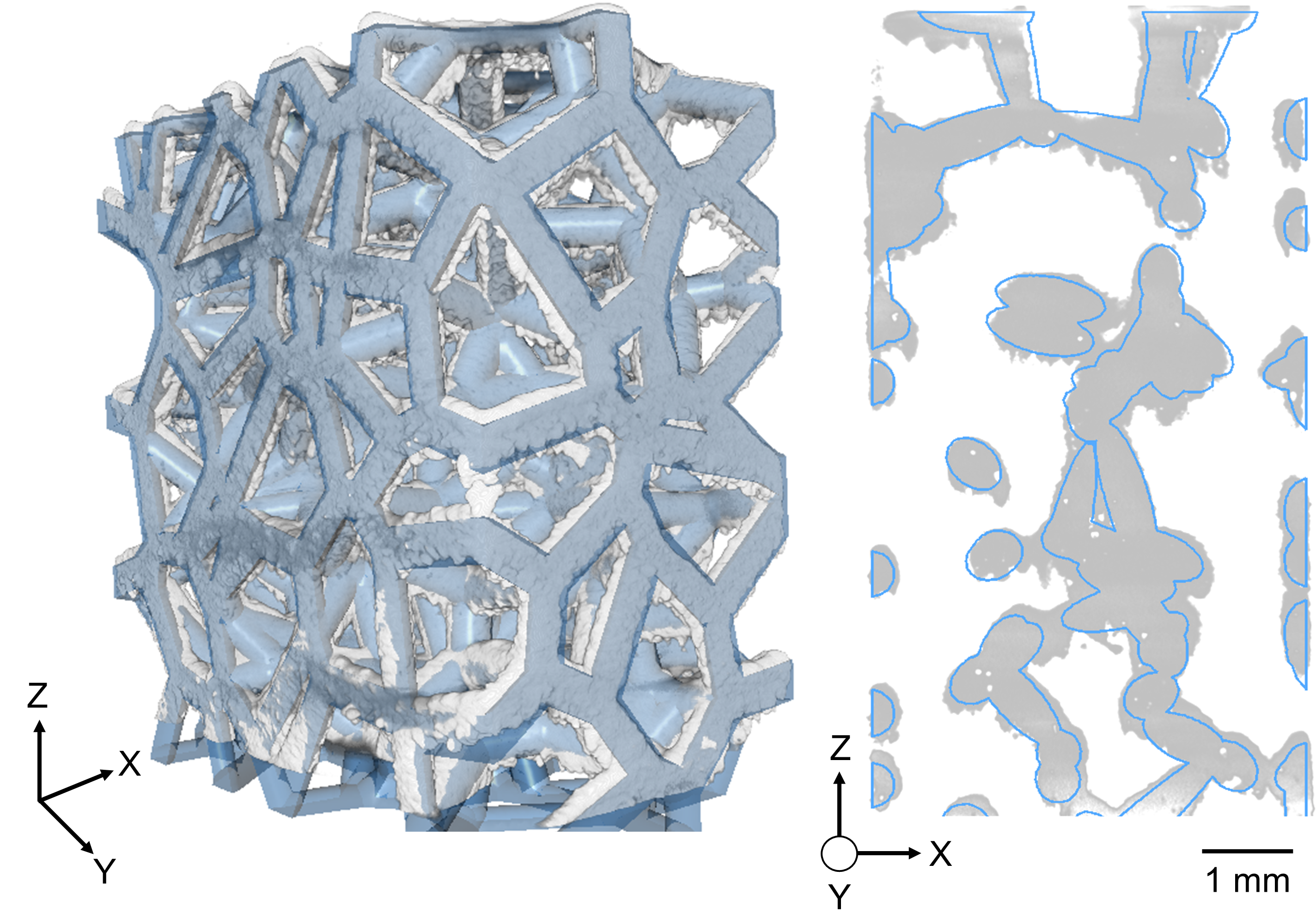}
    \caption{A 3D comparison of the XCT visual (shown in grey) and the original CAD (shown in blue) of a Ti-6Al-4V Voronoi lattice structure manufactured by laser powder bed fusion (left) and a 2D comparison of a slice along the XZ plane of the XCT visual (shown in grey) and the original CAD (shown in blue) of the lattice structure (right)}
    \label{fig:fidelity}
\end{figure}

In literature where dimensional inaccuracies have been reported, significant under-sizing and over-sizing of additively manufactured lattices compared to the original CAD are observed, sometimes over 100\%, as reviewed by Echeta et al \cite{echeta2020review}. Dimensional inaccuracies in AM lattices depend on numerous factors, including build file setup (location, orientation, recoater offset angle, etc.) \cite{moylan2013lessons}, PBF process parameters \cite{yasa2009investigation, patel2020towards}, CAD file resolution \cite{chougrani2017lattice}, powder size and morphology \cite{sames2016metallurgy, shanbhag2020effect}, and beam properties such as wavelength, operating mode, shape and quality  \cite{lee2017lasers, gong2012review}.

\hl{All types of lattice structures including strut-based, surface-based, plate-based, honeycombs, and stochastic lattices present their own unique setting of challenges for PBF manufacturing.} When it comes to strut-based lattices, higher dimensional inaccuracies are reported for diagonal and horizontal struts, when compared to vertically printed struts \cite{liu2017elastic, melancon2017mechanical, cuadrado2017influence}. This is because horizontal and diagonal struts contain the highest proportion of overhanging features which are known to have the highest surface roughness-related concerns, as noted in section \ref{roughness}. \hl{Additionally, distortion and excessive material accumulation has been previously reported at the nodes of strut-based lattices} \cite{gumruk2013static, leary2016selective, tancogne2016additively}, \hl{with comparatively smoother nodes observed in surface-based lattices} \cite{al2018topology}. \hl{Plate-based lattices also have issues during production using PBF such as overhanging features with zero inclination angle} \cite{liu2021mechanical}\hl{, and closed-cell topologies} \cite{crook2020plate} \hl{making the removal of powder from the lattices challenging after PBF. Honeycomb lattices are a class of lattice structures which would not have the same challenges as strut or surface based lattices during PBF manufacturing due to the extrusion of a 2D unit cell into a 3D lattice structure. As long as the 2D unit cell dimensions are within the feature resolution possible by PBF, honeycomb lattices should be relatively simple to manufacture. Lastly, stochastic lattices are the most challenging lattice type to manufacture using PBF. Stochastic lattice structures such as the Voronoi lattice structures shown in} Figure~\ref{fig:XCT} and Figure~\ref{fig:fidelity} \hl{provide control only over the strut thickness and the  spacing between the centroids of the Voronoi cells. This leads to numerous overhanging struts, many of which could have non self-supporting overhanging angles} $<$\ang{30} \hl{(with respect to the PBF build plate) thereby leading to the large deviation from CAD as shown in} Figure~\ref{fig:fidelity} \hl{and could lead to print failures at greater length scales. The spinodoid type of lattice structures developed by Kumar et al.} \cite{kumar2020inverse} \hl{provide additional control in the creation of stochastic lattice structures and might provide more opportunities in the area of printable stochastic lattice structures.}

For inclined up-skin surfaces in diagonal struts, a 'stair step' effect is commonly observed, wherein the edges of individual layers during PBF manufacturing are visible alongside effects of partially fused adhered powder \cite{strano2013surface, townsend2016surface}. This is caused primarily due to the stepped approximation of curves in inclined surfaces by layers for PBF, which further adds to dimensional inaccuracies of a manufactured lattice structure. In 2D and 3D comparison of a Voronoi lattice structure manufactured using LPBF shown in Figure \ref{fig:fidelity}, numerous dimensional anomalies of the printed lattice structure (compared to original CAD) can be observed, especially for the horizontal and diagonal struts which consist of down-skin surface areas. Additionally, nodes within strut based lattices are known to be regions with highest dimensional inaccuracies and defects \cite{echeta2020review}, especially when compared to surface based lattices \cite{al2018topology}. Nodes in strut based lattices are complex surfaces which involve a combination of up-skin, down-skin, and vertical surfaces. This complexity adds to the 'stair step' effect \cite{al2018topology} alongside effects of partially fused adhered powder \cite{li2016development}. Additionally, there could a compounded effect of joining multiple struts together in the subsequent layers, wherein due to residual stresses caused by large areas, the struts may not converge to a single point in space as observed in the CAD. These issues of geometric fidelity in AM lattices would be deleterious for bone replacement applications, and hence must be evaluated well before use.

X-ray computer tomography (XCT) of AM lattices is the most common method currently used to evaluate the geometric fidelity of printed lattices by comparison of the XCT scanned model with the original CAD model \cite{du2018x, vilardell2019topology, echeta2020review}, as also demonstrated in Figure \ref{fig:fidelity}. Other methods include scanning electron microscopy (SEM), optical microscopy, and Vernier calliper based comparisons of the AM lattice with the original CAD dimensions \cite{echeta2020review}.

\subsubsection{\hl{Advanced considerations for surface modifications}}

\hl{Porous titanium is unable to simulate organic components of bone. Surface modifications to porous titanium structures in post processing have shown to improve the biological performance and osseointegration} \cite{yavari2014bone, yan1997bonding, wang2008vitro, deering2021current, wang2016topological, wang2017analysis, asri2017corrosion, chouirfa2019review, de2013bone, park2011effect, xiu2016tailored, deering2020composite}. \hl{Surface modifications may include mechanical, chemical or physical methods} \cite{liu2004surface}. \hl{Mechanical methods such as machining, grinding, polishing and blasting aim to smooth or roughen surface topography to improve osteoblast adhesion} \cite{de2013bone}. \hl{Chemical surface modification methods such as alkaline treatments, hydrogen peroxide treatment, and biochemical modifications are typically aimed at improving biocompatibility, bioactivity or to induce a specific cell and tissue response} \cite{de2013bone,  park2011effect, xiu2016tailored, deering2020composite}. \hl{Physical modifications such as sprays, vapour and ion deposition modify surfaces to improve wear and corrosion resistance} \cite{liu2004surface}. \hl{One surface modification that has proven to improve osseointegration and increase osteogenesis is hydroxyapatite coatings} \cite{yadroitsava2019bone, park2011effect}.

\hl{Surface modifications may also reduce the risk of surface-bonded powder releasing into the body and causing tissue inflammation possibly due to macrophage activation} \cite{hollander2006structural}. \hl{Surface-bonded powder particles are also deleterious to an implants fatigue life as they act as stress concentrators} \cite{hrabe2011compression}. \hl{There have been notably few} \textit{in vivo} \hl{investigations into the effects of surface modification on titanium lattice structures produced the PBF AM technologies} \cite{deering2021current}. \hl{Future experimental work and systemic reviews should aim to quantify and discuss the effect of surface modification made to titanium lattice structures produced through PBF AM on osseointegration, osteogenesis and inflammatory tissue response.}

\subsubsection{Advanced considerations for trabecular geometry}
\begin{figure}[htbp]
    \centering
    \captionsetup{justification=centering}
    \includegraphics[width=15cm,keepaspectratio]{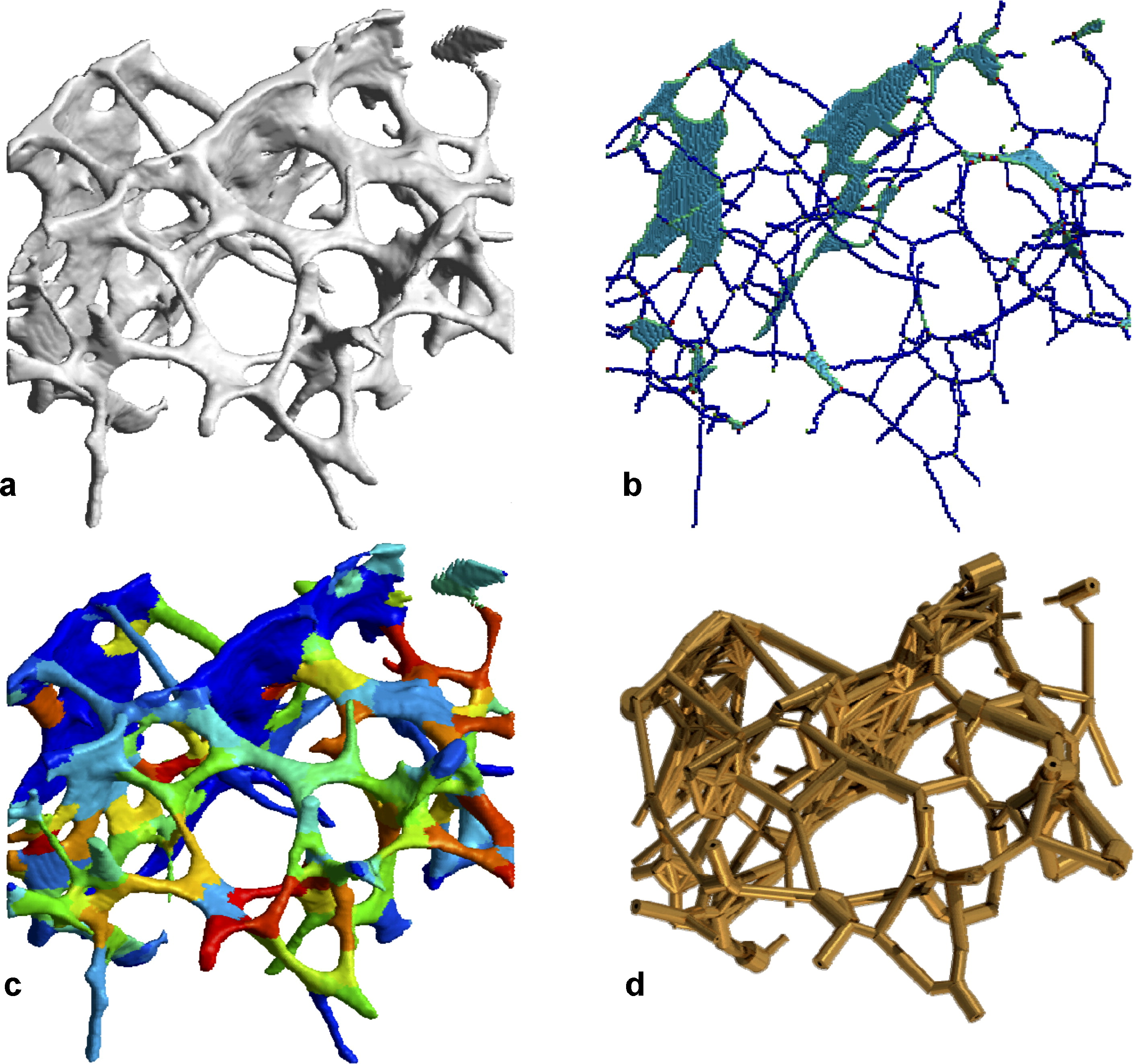}
    \caption{van Lenthe et al., used a micro computed tomographical (micro-CT) image  of a human trabecular bone to develop a specimen-specific beam finite-element model. The process is shown above: (a) micro-CT reconstruction, (b) point cloud generation, (c) multi-colour dilation, and (d) assignment of element thickness and volume to finite element beams. Original figure obtained from \cite{van2006specimen}}
    \label{fig:vanlenth}
\end{figure}

This review compared trabecular bone micro-architecture to lattice parameters that can be deployed in additive manufacturing. A simplistic model of trabeculae thickness and spacing was used to relate to the lattice parameters of strut thickness and pore size, respectively. However, trabecular bone has a more complex micro-architecture than traditional strut-based lattices and therefore surface-based lattices should be considered more closely in this application. Furthermore, van Lenthe et al recognized that trabecular bone exhibits both rod and plate-like behavior and suggested that a combined strut- and surface-based lattice configuration may best represent the trabecular bone micro-structure, see Figure \ref{fig:vanlenth} \cite{van2006specimen}; such structures can now be closer to attainable from a manufacturability stand point. 

Callens et al. proposed a mathematical model for describing local and global trabecular micro-architecture which suggested that bone does not exhibit mean zero curvature, like those exhibited in TMPS structures \cite{callens2020local}. Therefore, a stochastic surface-based lattice may be the best selection for modelling trabecular bone. This has been successfully attempted by Kumar et al. \cite{kumar2020inverse} and their work should be considered an emerging opportunity for the design of titanium lattice architectures for bone replacement and augmentation. Overall, surface lattices have shown success in replicating the mechanical properties of trabecular bone for titanium and titanium alloys and it is recommended that this avenue be pursued further \cite{yadroitsava2019bone}. 

\section{Conclusions}
This review covers the breath of human bone geometry, mechanical properties of cortical and trabecular bone and the attempts made at replacing these tissues with additively manufactured titanium lattice structures. Human bone, particularly trabecular bone varies significantly with age, sex, disease state, skeletal location and region of the individual bones. Therefore, a site specific and function specific design approach should be considered when designing for human bone replacement, repair and augmentation. Titanium lattice structures generated through additive manufacturing have shown success in replicating cortical bone mechanical properties, promoting osseointegration for improved implant fixation and reduction of stress shielding at the bone-implant interface.

Overall, many studies were able to tune lattice parameters to obtain a Young’s modulus and compression stiffness within the range of human cortical bone, with challenges in addressing the porous network architecture concomitantly. Cortical bone is roughly 5-15\% porosity and titanium lattices with roughly 50-70\% porosity were most successful in achieving comparable stiffness and compressive strength. Matching mechanical properties of trabecular bone was less achievable in the current literature. Trabecular thickness, or feature thickness, of human trabecular bone is on the cusp of what is attainable with existing metal additive manufacturing powder bed fusion technologies. Additionally, matching trabecular thickness to feature thickness in titanium-based lattices does not account for the difference in bulk modulus. Therefore, matching feature thickness to trabecular bone thickness is not a recommended technique for matching the mechanical properties of titanium lattice structures to those of human trabecular bone. Control of pore size, porosity and lattice type may yield better results when attempting to replace trabecular bone with additively manufactured titanium lattices. 

\hl{Young's modulus and compressive strength can be correlated to porosity of cellular latticed materials through the Gibson-Ashby model} \cite{gibson2003cellular}. \hl{The Gibson-Ashby model has been used to predict compressive mechanical properties of Ti-6Al-4V lattice structures by adjusting lattice type, feature thickness and pore size to control porosity} \cite{benedetti2019study, maconachie2019slm}. \hl{Future efforts should build upon this work to develop a more robust transfer function that better relates the bulk moduli of titanium and Ti-6Al-4V to the anisotropic and variable lattice parameters required to generate structures that more seamlessly match the mechanical properties of site and patient specific cortical and trabecular bone}. \hl{Future work should provide a comparison of the apparent density of dry and wet bone to better characterize human bone material.} Due to the large variations in bone porosity and microstructure throughout the human skeleton and the increase of bone porosity with age and in females, patient specific design may yield the best outcome with respect matching mechanical properties of bone with additively manufactured lattices. Improved lattice design for bone replacement and augmentation will allow for improved orthopaedic implant design and may ultimately reduce the risk of stress shielding at the bone implant interface.

\section{Acknowledgements}
Martine McGregor is supported by the Natural Sciences and Engineering Research Council of Canada doctoral scholarship. Sagar Patel and Mihaela Vlasea appreciate the funding support received from Federal Economic Development Agency for Southern Ontario (FedDev Ontario grant number 814654). Additionally, Sagar Patel and Mihaela Vlasea would like to acknowledge the help of Jerry Ratthapakdee and Henry Ma with the deployment and characterization of the laser powder bed fusion builds. The authors would like to thank nTopology, Inc. for supporting us with a license to their software for the lattice visualizations.

\bibliography{references.bib}

\end{document}